%% file: paper.tex
\documentclass[sigconf]{acmart}
\usepackage{standalone}
\usepackage[english]{babel}
\usepackage{multirow}
\usepackage{tikz}
\usepackage[section]{placeins} 
\usetikzlibrary{positioning, arrows.meta, shapes.geometric, calc, backgrounds, patterns, fit, shapes.misc, angles, quotes}

\allowdisplaybreaks[2]

\definecolor{riskRed}{RGB}{0, 0, 0}               
\definecolor{riskOrange}{RGB}{0, 114, 178}        
\definecolor{riskGreen}{RGB}{230, 159, 0}         
\definecolor{flowS1}{RGB}{0, 114, 178}            
\definecolor{flowS2}{RGB}{213, 94, 0}             
\definecolor{netblue}{RGB}{65, 105, 225}
\definecolor{procgreen}{RGB}{50, 205, 50}
\definecolor{lossred}{RGB}{220, 20, 60}
\definecolor{datagray}{RGB}{128, 128, 128}
\renewcommand\footnotetextcopyrightpermission[1]{} 
\setcopyright{none}

\acmDOI{}

\acmISBN{}

\acmPrice{}

\begin{document}
\title{NeuroRisk: Physics-Informed Neural Optimization for Risk-Aware Traffic Engineering}

\author{%
\begingroup
\fontsize{10.5pt}{12.5pt}\selectfont
\textnormal{
Yingming Mao\textsuperscript{1,5},
Ximeng Liu\textsuperscript{2},
Jingyi Cheng\textsuperscript{2},
Xiyuan Liu\textsuperscript{3},
Jiashuai Liu\textsuperscript{1},
Yike Liu\textsuperscript{1},
Zhen Yao\textsuperscript{4},
Yuzhou Zhou\textsuperscript{1},
Siyuan Feng\textsuperscript{5},
Qiaozhu Zhai\textsuperscript{1},
Shizhen Zhao\textsuperscript{2}
}\\[0.45em]
\textnormal{
\textsuperscript{1}Xi'an Jiaotong University \quad
\textsuperscript{2}Shanghai Jiao Tong University \quad
\textsuperscript{3}Nanyang Technological University
}\\
\textnormal{
\textsuperscript{4}Huawei\quad
\textsuperscript{5}Shanghai Innovation Institute
}
\endgroup
}


\renewcommand{\shortauthors}{Mao et al.}

\begin{abstract}
In production Wide-Area Networks (WANs), correlated failures dominate availability losses, forcing operators to reserve large safety margins that leave substantial capacity underutilized. Achieving high utilization under strict availability targets therefore requires risk-aware Traffic Engineering (TE) over dozens to hundreds of probabilistic failure scenarios—yet solving this problem at operational timescales remains elusive. We demonstrate that existing risk-aware formulations can be unified under an embedded Sort-and-Select structure, exposing a fundamental trade-off between expressiveness and tractability: classical optimizers either restrict scenario selection for efficiency or incur prohibitive decomposition costs.
While deep learning appears promising, prior Deep TE methods mainly target maximum link utilization and rely on scaling-based feasibility, which fundamentally breaks under explicit capacity constraints and scenario-dependent risk. We present NeuroRisk, a physics-informed deep unrolled optimizer that exploits the structure of Sort-and-Select. NeuroRisk enforces feasibility via gated edge-local reservations and represents scenario sets through permutation-invariant, gradient-aligned cues.
Evaluations on production-style WANs show that NeuroRisk achieves small optimality gaps relative to the solver with orders of magnitude speedup $(10^2- 10^5 \times)$ on risk objectives, while outperforming neural baselines on nominal throughput.

\end{abstract}

\maketitle

\input{body}

\end{document}

%% file: body.tex
\section{Introduction}
\label{sec:intro}

In Wide-Area Networks (WANs), failures are unavoidable. While operators could overprovision capacity to tolerate such events, the high cost of bandwidth makes this approach economically prohibitive~\cite{b4_after,offenfailure2}. Instead, the industry relies on Traffic Engineering (TE)~\cite{AKYILDIZ20141,alizadehCONGADistributedCongestionaware2014,bensonMicroTEFineGrained2011,832225,swan,10.1145/2486001.2486019,10.1145/1555349.1555377,10.1145/1090191.1080122} to maximize utilization in the presence of failures. However, traditional reactive TE--which re-optimizes flows after a failure is detected--suffers from inevitable control-loop latencies, leading to transient overloads~\cite{ncflow, ssdo}. This has driven a shift toward proactive TE, which prepares the network for uncertainty by optimizing traffic distributions for potential failure scenarios in advance.

Existing proactive TE methods model uncertainty in two main ways. The first group focuses on worst-case guarantees (e.g., FFC~\cite{ffc}), ensuring bandwidth-guaranteed operation under any predefined failure, typically at the expense of capacity over-reservation. The second group employs probabilistic models: TeaVaR~\cite{teavar} optimizes tail risk (CVaR~\cite{var}), while FloMore~\cite{flomore} supports per-flow quantile constraints--albeit at the cost of significantly higher computational complexity through iterative decomposition.

These seemingly disparate methods share a common structure that \textbf{each induces a decision-dependent ordering over failure scenarios and then selects a subset to optimize}, which we term the \emph{Sort-and-Select} paradigm. FFC selects only the top-ranked scenario; CVaR selects the tail; FloMore selects a single rank per flow. This abstraction reveals the root cause of the expressiveness--tractability trade-off: {embedding sorting inside optimization introduces combinatorial structure}. Using specific mask types (e.g., CVaR) restores tractability by convexifying the objective, but sacrifices flexibility; allowing per-flow masks captures richer preferences but explodes complexity. As the number of scenarios grows, traditional solvers struggle to run in real-time.

For accelerating combinatorial optimization~\cite{co}, deep learning has emerged as a powerful tool~\cite{BENGIO2021405,10.1093/nsr/nwae132,pinnforode,pinneik,gaussianpinn,pointer,neuralCOwithRL,dai2018,Goodfellow-et-al-2016,CHEN1995915}. Instead of using exact algorithms to search for optimal solutions, this paradigm treats optimization as a mapping from problem context to decision variables, and trains a neural network to approximate this mapping. At inference time, the complex search process is reduced to a simple forward pass, drastically cutting down the computation time. This approach has been widely adopted in Traffic Engineering--prior work has learned demand-to-routing mappings (Figret~\cite{figret}), generalized across topologies (Geminet~\cite{liu2025geminetlearningdualitybasediterative}), and incorporated per-link failure probabilities (FauTE~\cite{faute}). However, applying these techniques to risk-aware TE faces two fundamental barriers.

The first is the \emph{Feasibility Barrier}. Existing Deep TE methods have been largely developed to minimize maximum link utilization (MLU), whose objective is fundamentally misaligned with failure-aware TE. MLU measures network congestion rather than the allocation of limited capacity under resource scarcity. Failure-aware formulations shift from implicit congestion objectives to explicit capacity constraints. Existing approaches claim that post-hoc scaling can generalize to this setting~\cite{dote}; we find this assumption invalid, leading to collateral damage and systematic short-path bias.

The second is the \emph{Representation Barrier}. Deep TE methods have progressively increased input complexity: from simple demand matrices~\cite{dote} to concatenated vectors of demands and link failure rates~\cite{faute}, and recently adopting Graph Neural Networks to capture topology features~\cite{harp}. Yet, none of the methods encode the full probabilistic scenario set $\mathcal{Q}$ required by risk-aware TE: correlated multi-link failures $q \in \mathcal{Q}$ with probabilities $p_q$, where both the size and composition of $\mathcal{Q}$ change at runtime. FauTE's soft penalty optimizes a surrogate rather than true risk objectives, and naive padding/aggregation erases scenario structure.

To address these barriers, we propose NeuroRisk, a physics-informed deep unrolled optimizer for scenario-dependent \emph{Sort-and-Select} objectives. Crucially, the structure of these objectives allows the gradient computation to commute with scenario aggregation, naturally motivating a gradient-driven unrolled optimizer rather than a black-box predictor~\cite{BENGIO2021405,10.1093/nsr/nwae132,dote}. For feasibility and optimization, we introduce \emph{Gated Reservation}, which transforms the decision space from globally-coupled flow splits to edge-local reservation ratios, guaranteeing capacity feasibility by construction and enabling stable, high-quality capacity allocation. For representation, a physics engine projects $\mathcal{Q}$ into permutation-invariant gradient cues, preserving scenario-dependent structure. Our formulation subsumes deterministic TE ($|\mathcal{Q}|=1$), making NeuroRisk a general-purpose Deep TE optimizer. 

Our key contributions are:
\begin{itemize}
    \item \emph{Sort-and-Select Risk Modeling:} We formalize a unified \emph{Sort-and-Select} view of risk-aware TE, exposing the root cause of the expressiveness--tractability trade-off in existing optimizers and providing operators a clean knob to express scenario-dependent risk preferences.

    \item \emph{Feasibility-Preserving Reparameterization:} Scaling is fundamentally flawed for risk-aware objectives. Instead, Gated Reservation reparameterizes routing into edge-local reservations with tunnel-level gating, guaranteeing capacity feasibility by construction.

    \item \emph{Structure-Aware Scenario Representation:} We encode probabilistic failure scenarios through physics-derived gradient cues aligned with \emph{Sort-and-Select} objectives, enabling generalization across variable $\mathcal{Q}$, topologies, and tunnel sets while matching solver-quality solutions with orders-of-magnitude speedups.
\end{itemize}

\section{Background and Motivation}
\label{sec:background}

This section introduces the risk-aware TE setting and illustrates why existing formulations lead to a computational bottleneck.
We route flows $f \in \mathcal{F}$ with demand $D_f$ over candidate tunnels $\mathcal{T}_f$ using split ratios $\mathbf{x} = \{x_{f,t}\}$, where $x_{f,t} \ge 0$; consistent with prior work~\cite{ffc}, we allow $\sum_t x_{f,t} > 1$ (over-provisioning for failover) or $< 1$ (under-provisioning under capacity scarcity).
Uncertainty is captured by scenario set $\mathcal{Q}$, which includes the no-failure (nominal) scenario: each scenario $q$ occurs with probability $p_q$ ($\sum_{q \in \mathcal{Q}} p_q = 1$) and disables tunnels via the binary indicator $\alpha_{t,q} \in \{0,1\}$ (1 if tunnel $t$ survives in $q$, 0 otherwise), inducing per-flow loss $\ell_{f,q}$—the fraction of demand undelivered by surviving tunnels. Full notation is in Appendix~\ref{app:notation}.

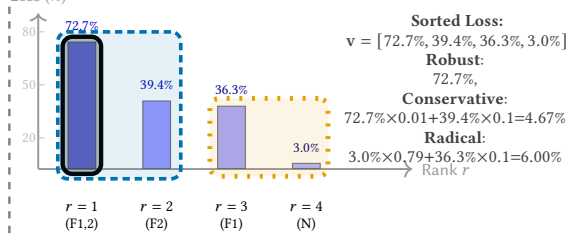
\begin{figure*}[htbp]
    \centering 

    \resizebox{1\textwidth}{!}{
        \begin{tikzpicture}[font=\sffamily, scale=0.9]

\tikzset{
    netnode/.style={circle, draw=black!80, thick, fill=white, minimum size=0.65cm, inner sep=1pt, font=\small\bfseries},
    link/.style={->, >=Latex, line width=1.2pt, gray!60},
    link_label/.style={font=\scriptsize, fill=white, inner sep=0.5pt, opacity=0.9},
    failureX/.style={cross out, draw=red, line width=2pt, minimum size=8pt},
    bar_axis/.style={->, thick, gray!80},
    loss_bar/.style={draw=black!60, fill=gray!30, anchor=south, minimum width=0.4cm},
    bar_label/.style={font=\tiny, below=2pt},
    mask_box/.style={dashed, line width=1.2pt, rounded corners=4pt, fill opacity=0.1},
    robust_mask/.style={mask_box, draw=riskRed, fill=riskRed, fill opacity=0.25, solid, line width=1.8pt},
    conservative_mask/.style={mask_box, draw=riskOrange, fill=riskOrange, densely dashed, line width=1.5pt},
    radical_mask/.style={mask_box, draw=riskGreen, fill=riskGreen, loosely dotted, line width=2pt},
    section_title/.style={font=\large\bfseries, align=center, blue!40!black},
    subsection_title/.style={font=\bfseries, align=left, gray!30!black, anchor=west},
    demand_label/.style={font=\scriptsize\bfseries, text=blue!60!black, align=right}
}

\begin{scope}[local bounding box=col1]
    \node[section_title] at (1, 9.8) {1. Context: Topology \& Demands};

    \node[subsection_title] at (-2, 8.8) {Base Topology \& Total Demands};
    \node[netnode] (S1) at (0, 7.5) {S1};
    \node[netnode] (S2) at (0, 6.0) {S2};
    \node[netnode] (M) at (2, 6.75) {M};
    \node[netnode] (D) at (4, 6.75) {D};

    \node[demand_label, left=0.1cm of S1] {$D_{S_1 D}=15$G};
    \node[demand_label, left=0.02cm of S2] {$D_{S_2 D}=12.5$G};

    \draw[link] (S1) -- node[link_label, right, pos=0.3] {10G} (M);
    \draw[link] (S2) -- node[link_label, right, pos=0.3] {10G} (M);
    \draw[link] (M) -- node[link_label, above] {7.5G} (D);
    \draw[link] (S1) to[bend left=40] node[link_label, above] {10G (Dir)} (D);
    \draw[link] (S2) to[bend right=40] node[link_label, below] {10G (Dir)} (D);

    \node[subsection_title] at (-2, 4.8) {Failure Scenarios Spectrum};
    
    \tikzset{mini_net/.style={scale=0.35, every node/.style={scale=0.7}}}

    \node[anchor=west, font=\scriptsize] at (-2, 3.8) {\textbf{N. No Failure. $p_{N}=0.79$}};
    \begin{scope}[shift={(2.5, 3.8)}, mini_net]
        \node[netnode] (s1) at (0,1) {}; \node[netnode] (s2) at (0,-1) {}; \node[netnode] (d) at (3,0) {};
        \draw[link, thin] (s1) to[bend left] (d); \draw[link, thin] (s2) to[bend right] (d);
        \node[font=\tiny, black!60] at (1.5, -1.5) {All OK};
    \end{scope}

    \node[anchor=west, font=\scriptsize] at (-2, 2.6) {\textbf{F1. S1-D Direct Fails. $p_{F1}=0.1$}};
    \begin{scope}[shift={(2.5, 2.6)}, mini_net]
        \node[netnode] (s1) at (0,1) {}; \node[netnode] (s2) at (0,-1) {}; \node[netnode] (d) at (3,0) {};
        \draw[link, thin, gray!30] (s1) to[bend left] coordinate[midway](m1) (d); \draw[link, thin] (s2) to[bend right] (d);
        \node[failureX, scale=0.5] at (m1) {};
    \end{scope}

    \node[anchor=west, font=\scriptsize] at (-2, 1.4) {\textbf{F2. S2-D Direct Fails. $p_{F2}=0.1$}};
    \begin{scope}[shift={(2.5, 1.4)}, mini_net]
        \node[netnode] (s1) at (0,1) {}; \node[netnode] (s2) at (0,-1) {}; \node[netnode] (d) at (3,0) {};
        \draw[link, thin] (s1) to[bend left] (d); \draw[link, thin, gray!30] (s2) to[bend right] coordinate[midway](m2) (d);
        \node[failureX, scale=0.5] at (m2) {};
    \end{scope}

    \node[anchor=west, font=\scriptsize] at (-2, 0.2) {\textbf{F1,2. Both Direct Fail. $p_{F1,2}=0.01$}};
    \begin{scope}[shift={(2.5, 0.2)}, mini_net]
        \node[netnode] (s1) at (0,1) {}; \node[netnode] (s2) at (0,-1) {}; \node[netnode] (d) at (3,0) {};
        \draw[link, thin, gray!30] (s1) to[bend left] coordinate[midway](m1) (d); \draw[link, thin, gray!30] (s2) to[bend right] coordinate[midway](m2) (d);
        \node[failureX, scale=0.5] at (m1) {}; \node[failureX, scale=0.5] at (m2) {};
    \end{scope}

\end{scope}
\draw[gray, dashed, line width=1pt] ($(col1.north east)+(0,0)$) -- ($(col1.south east)+(0,0)$);

\begin{scope}[shift={(4.6, 0)}, local bounding box=col2]
    \node[section_title] at (2.5, 9.8) {2. Strategies: Allocations $\mathbf{x}$};
    \node[font=\scriptsize, gray!80!black, align=center] at (2.6, 9.2) {Meeting Demands: $D_{S_1D}=15$G, $D_{S_2D}=12.5$G\\(Visualized under Normal Scenario)};

    \node[subsection_title] at (0, 8.2) {Decision 1: Ignore Failure.};

    \begin{scope}[shift={(0.5, 5.0)}]
        \node[netnode] (S1_d1) at (0, 1.5) {S1}; \node[netnode] (S2_d1) at (0, 0) {S2};
        \node[netnode] (M_d1) at (2, 0.75) {M}; \node[netnode] (D_d1) at (4, 0.75) {D};
        
        \draw[->, line width=1.5pt, flowS1] (S1_d1) to[bend left=40] node[midway, above, link_label, text=flowS1] {10G} (D_d1);
        \draw[->, line width=1.5pt, flowS1] (S1_d1) -- node[above, sloped, link_label, text=flowS1] {5G} (M_d1);
        
        \draw[->, line width=1.5pt, flowS2, densely dashed] (S2_d1) to[bend right=40] node[midway, below, link_label, text=flowS2] {10G} (D_d1);
        \draw[->, line width=1.5pt, flowS2, densely dashed] (S2_d1) -- node[below, sloped, link_label, text=flowS2] {2.5G} (M_d1);
        
        \draw[->, line width=1.5pt, flowS1] (M_d1) to[bend left=10] node[above, link_label, text=flowS1] {5G} (D_d1);
        \draw[->, line width=1.5pt, flowS2, densely dashed] (M_d1) to[bend right=10] node[below, link_label, text=flowS2] {2.5G} (D_d1);

    \end{scope}

    \begin{scope}[shift={(0.5, -0.5)}]
        \node[subsection_title] at (-0.5, 4.0) {Decision 2: Prevent Failure};

        \begin{scope}[shift={(0, 0.8)}]
            \node[netnode] (S1_d2) at (0, 1.5) {S1}; \node[netnode] (S2_d2) at (0, 0) {S2};
            \node[netnode] (M_d2) at (2, 0.75) {M}; \node[netnode] (D_d2) at (4, 0.75) {D};
    \draw[black!60, dashed, line width=1pt] 
        ($(S1_d2.north west)+(-0.2,-0.7)$) -- 
        ($(S1_d2.north east)+(0.2,-0.7)$) -- 
        ($(S1_d2.south east)+(0.2,0.7)$) -- 
        ($(S1_d2.south west)+(-0.2,0.7)$) -- 
        cycle; 

        \node[font=\small, black!60] 
        at ($(S2_d2.east)+(1.7,2.5)$) 
        {Excess allocated bandwidth sits idle.};
            \draw[->, line width=1.5pt, flowS1] (S1_d2) to[bend left=40] node[midway, above, link_label, text=flowS1] {10G} (D_d2);
            \draw[->, line width=1.5pt, flowS1] (S1_d2) -- node[above, sloped, link_label, text=flowS1] {4.1G} (M_d2);
            
            \draw[->, line width=1.5pt, flowS2, densely dashed] (S2_d2) to[bend right=40] node[midway, below, font=\scriptsize, black, fill=white, opacity=0.9, text=flowS2] {10G} (D_d2);
            \draw[->, line width=1.5pt, flowS2, densely dashed] (S2_d2) -- node[below, sloped, link_label, text=flowS2] {3.4G} (M_d2);
            
            \draw[->, line width=1.5pt, flowS1] (M_d2) to[bend left=10] node[above, link_label, text=flowS1] {4.1G} (D_d2);
            \draw[->, line width=1.5pt, flowS2, densely dashed] (M_d2) to[bend right=10] node[below, link_label, text=flowS2] {3.4G} (D_d2);

        \end{scope}
    \end{scope}
\end{scope}
\draw[gray, dashed, line width=1pt] ($(col2.north east)+(0.0,0)$) -- ($(col2.south east)+(0.00,0)$);

\begin{scope}[shift={(10, 0)}, local bounding box=col3]
    \node[section_title] at (3.5, 9.8) {3. Analysis: Sorted Loss \& Selection};

    \begin{scope}[shift={(0.5, 8.5)}]
        \draw[robust_mask, fill opacity=0.3] (0,0.4) rectangle (0.5,0.9); \node[anchor=west, font=\scriptsize] at (0.6, 0.65) {\textbf{Robust}};
        \draw[conservative_mask, fill opacity=0.3] (2.8,0.4) rectangle (3.3,0.9); \node[anchor=west, font=\scriptsize] at (3.4, 0.65) {\textbf{Conservative} };
        \draw[radical_mask, fill opacity=0.3] (5.8,0.4) rectangle (6.3,0.9); \node[anchor=west, font=\scriptsize] at (6.4, 0.65) {\textbf{Radical}};
        
    \end{scope}

    \begin{scope}[shift={(0, 5.0)}]
        \node[subsection_title, anchor=west] at (0, 3.2) {Analysis of Decision 1};
        \node[font=\scriptsize, anchor=west, gray!50!black, align=center] at (5.3, 2.2) {\textbf{Sorted Loss:}\\ $\mathbf{v} = [73.3\%, 40.0\%, 33.3\%, 0\%]$};
        
        \draw[bar_axis] (0.5,0) -- (6.5,0) node[right] {\scriptsize Rank $r$};
        \draw[bar_axis] (0.5,0) -- (0.5,2.5) node[above] {\scriptsize Loss (\%)};
        \foreach \y/\label in {0.5/20, 1.35/50, 2.2/80} \draw[gray!50, thin] (0.5, \y) -- (0.6, \y) node[left, font=\tiny] {\label};

\node[loss_bar, minimum height=1.825cm, fill=flowS2!60] (d1_r0) at (1.2,0) {}; 
\node[bar_label, align=center, font=\tiny] at (1.2,-0.3) {$r=1$\\(F1,2)};
\node[font=\tiny, above, flowS2!80!black, yshift=1pt] at (d1_r0.north) {73.3\%};

\node[loss_bar, minimum height=1.0cm, fill=flowS2!50] (d1_r1) at (2.4,0) {}; 
\node[bar_label, align=center, font=\tiny] at (2.4,-0.3) {$r=2$\\(F2)};
\node[font=\tiny, above, flowS2!70!black, yshift=1pt] at (d1_r1.north) {40.0\%};

\node[loss_bar, minimum height=0.825cm, fill=flowS2!40] (d1_r2) at (3.6,0) {}; 
\node[bar_label, align=center, font=\tiny] at (3.6,-0.3) {$r=3$\\(F1)};
\node[font=\tiny, above, flowS2!60!black, yshift=1pt] at (d1_r2.north) {33.3\%};

\node[loss_bar, minimum height=0.02cm, fill=gray!20, inner sep=0pt] (d1_r3) at (4.8,0) {}; 
\node[bar_label, align=center, font=\tiny] at (4.8,-0.3) {$r=4$\\(N)};
\node[font=\tiny, above, black!60] at (d1_r3.north) {0\%};      
        \node[robust_mask, fit=(d1_r0)(d1_r0), inner sep=2pt] (mask_r_d1) {}; 
        \node[conservative_mask, fit=(d1_r0)(d1_r1), inner sep=4pt] (mask_c_d1) {}; 
        \node[radical_mask, fit=(d1_r2)(d1_r3), inner sep=3pt] (mask_rad_d1) {}; 
        
        \node[below, font=\scriptsize, gray!50!black, align=center] at ($(mask_rad_d1.east)+(2,1.5)$) {\textbf{Robust}:\\$73.3\%$,\\\textbf{Conservative}:\\$73.3\%{\times}0.01{+}40.0\%{\times}0.1{=}4.73\%$\\\textbf{Radical}:\\$0\%{\times}0.79{+}33.3\%{\times}0.1{=}3.33\%$};
    \end{scope}

    \begin{scope}[shift={(0, 0.5)}]
        \node[subsection_title, anchor=west] at (0, 3) {Analysis of Decision 2};
         \node[font=\scriptsize, anchor=west, gray!50!black, align=center] at (5.3, 2.2) {\textbf{Sorted Loss:}\\$\mathbf{v} = [72.7\%, 39.4\%, 36.3\%, 3.0\%]$};
        
        \draw[bar_axis] (0.5,0) -- (6.5,0) node[right] {\scriptsize Rank $r$};
        \draw[bar_axis] (0.5,0) -- (0.5,2.5) node[above] {\scriptsize Loss (\%)};
        \foreach \y/\label in {0.5/20, 1.35/50, 2.2/80} \draw[gray!50, thin] (0.5, \y) -- (0.6, \y) node[left, font=\tiny] {\label};
        
\node[loss_bar, minimum height=1.825cm, fill=blue!50] (d2_r0) at (1.2,0) {}; 
\node[bar_label, align=center, font=\tiny] at (1.2,-0.3) {$r=1$\\(F1,2)};
\node[font=\tiny, above, blue!80!black, yshift=1pt] at (d2_r0.north) {72.7\%};

\node[loss_bar, minimum height=0.975cm, fill=blue!40] (d2_r1) at (2.4,0) {}; 
\node[bar_label, align=center, font=\tiny] at (2.4,-0.3) {$r=2$\\(F2)};
\node[font=\tiny, above, blue!70!black, yshift=1pt] at (d2_r1.north) {39.4\%};

\node[loss_bar, minimum height=0.9cm, fill=blue!35] (d2_r2) at (3.6,0) {}; 
\node[bar_label, align=center, font=\tiny] at (3.6,-0.3) {$r=3$\\(F1)};
\node[font=\tiny, above, blue!60!black, yshift=1pt] at (d2_r2.north) {36.3\%};

\node[loss_bar, minimum height=0.075cm, fill=blue!30, inner sep=0pt] (d2_r3) at (4.8,0) {}; 
\node[bar_label, align=center, font=\tiny] at (4.8,-0.3) {$r=4$\\(N)};
\node[font=\tiny, above, blue!40!black, yshift=1pt] at (d2_r3.north) {3.0\%};

        \node[robust_mask, fit=(d2_r0)(d2_r0), inner sep=2pt] (mask_r_d2) {}; 
        \node[conservative_mask, fit=(d2_r0)(d2_r1), inner sep=4pt] (mask_c_d2) {}; 
        \node[radical_mask, fit=(d2_r2)(d2_r3), inner sep=3pt] (mask_rad_d2) {}; 
        \node[below, font=\scriptsize, gray!50!black, align=center] at ($(mask_rad_d2.east)+(2,1.5)$) {\textbf{Robust}:\\$72.7\%$,\\\textbf{Conservative}:\\$72.7\%{\times}0.01{+}39.4\%{\times}0.1{=}4.67\%$\\\textbf{Radical}:\\$3.0\%{\times}0.79{+}36.3\%{\times}0.1{=}6.00\%$};

    \end{scope}
\end{scope}

\end{tikzpicture}
    }
    \caption{\textbf{A Unified Analysis Perspective for Risk-Aware TE Strategies.}
    \textbf{(Left)} A topology with two demands ($S1, S2 \to D$). Each demand has a choice between a risky direct tunnel (susceptible to failure) and a safe indirect tunnel (via $M$). The failure scenarios and their probabilities are listed below.
\textbf{(Middle)} Two distinct allocation strategies: Decision 1 (Radical) maximizes throughput by utilizing risky tunnels, assuming the normal scenario (prob. 0.79); Decision 2 (Robust) proactively throttles traffic and uses safe tunnels to survive failures.
    \textbf{(Right)} The \textit{Sorted Loss Analysis} reveals that these strategies--and existing paradigms like FFC (Robust) or TeaVaR (Conservative)--are essentially different ``selection masks'' prioritizing different ranks in the loss distribution.
    Sorted loss is the average flow incompletion: e.g., for F1,2 under Decision 1, $\ell_{S1} = 1 - 5/15 = 66.7\%$, $\ell_{S2} = 1 - 2.5/12.5 = 80.0\%$, average $= 73.3\%$; under Decision 2, $\ell_{S1} = 1 - 4.1/15 = 72.7\%$, $\ell_{S2} = 1 - 3.4/12.5 = 72.8\%$, average $= 72.7\%$.}
    \label{fig:risk_framework_analysis}

\end{figure*}

\subsection{An Illustration of Sorted-Loss Paradigm}
\label{subsec:risk_anatomy}

Risk-aware Traffic Engineering (TE) fundamentally deals with uncertainty. Operators must make routing decisions $\mathbf{x}$ before knowing which failure scenario $q \in \mathcal{Q}$ will materialize.
To illustrate the challenges in risk-aware TE, consider the setup in the left column of Figure \ref{fig:risk_framework_analysis}.
The network serves two demands ($D_{S_1D}=15$G, $D_{S_2D}=12.5$G) destined for $D$. Each source has two tunneling options: a high-capacity but \emph{risky direct tunnel} (e.g., $S1 \to D$, prone to failure with probability 0.1) and a \emph{safe indirect tunnel} via node $M$ ($S1 \to M \to D$).

We present two bandwidth allocation decisions to demonstrate how different risk preferences dictate the optimal choice (see middle column of Figure~\ref{fig:risk_framework_analysis}).
\emph{Decision 1} fully utilizes the direct tunnels, allocating the remaining demand to the safe tunnel via $M$.
\emph{Decision 2} shifts more traffic to the safe tunnel, providing redundancy at the cost of slightly under-serving $S1$ in the nominal scenario.
Unless otherwise specified, we evaluate decisions based on the flow incompletion rate, where lower values are preferred. We define the loss ratio of flow $f$ in scenario $q$ as defined in Equation~\eqref{eq:l_defination}:
\begin{equation}
    \ell_{f,q} = \max\left\{0, 1 - \sum_{t \in \mathcal{T}_f} x_{f,t} \cdot \alpha_{t,q}\right\}
    \label{eq:l_defination}
\end{equation}
\noindent Here $\ell_{f,q}\in[0,1]$ measures the undelivered fraction due to failures ($\alpha_{t,q}=0$). Notation is summarized in Appendix~\ref{app:notation}.
For exposition, we define scenario-level loss as the average over flows: $\ell_q = \frac{1}{|\mathcal{F}|}\sum_f \ell_{f,q}$.
Following prior work~\cite{prete}, we adopt this loss definition, with the framework agnostic to it.

\noindent\textbf{Unifying Diverse Strategies.}
The right column of Figure~\ref{fig:risk_framework_analysis} shows that these strategies share one structure: first \emph{sort} scenarios by loss (rank $r$), then \emph{select} a subset via a mask.  The \emph{Robust} view uses only the worst rank (dark mask); the \emph{Conservative} view uses the tail--the few highest-loss ranks (blue mask); the \emph{Radical} view uses the nominal and lower-loss ranks (amber mask). Thus ``risk preference'' is exactly a choice of \emph{selection mask} over the sorted loss distribution.

Applying these masks to the example shows how risk preference dictates allocation. Under the Robust view (worst-case only), Decision 2 is slightly better due to its more balanced indirect-tunnel allocation. The Conservative strategy, by aggregating over the tail, also favors Decision 2; the Radical strategy, by emphasizing nominal and lower-loss scenarios, clearly favors Decision 1.
Figure~\ref{fig:risk_framework_analysis} summarizes this \emph{Sort-and-Select} view. We next formalize it as a unified optimization problem and identify computational bottlenecks.

\subsection{Unified Mathematical Formulation}
\label{sec:unified_formulation}

We now view risk-aware TE through a different lens: rather than treating CVaR, chance constraints, and robust TE as separate models, we explicitly model them as the \emph{same} optimization problem with a tunable \emph{selection mask} over the sorted-loss ranks (Figure \ref{fig:risk_framework_analysis}).
This paradigm shift decouples risk preference (which ranks you care about) from physics (capacity and routing), turning previously rigid, objective-specific formulations into a single, expressive template.
Crucially, it also makes the real bottleneck unambiguous: the only nontrivial difficulty is enforcing sorting inside the optimization, which is exactly where prior approaches compromise (by specific masks or aggregating scenarios) and where our solver later removes the complexity.

Unlike the scenario-level illustration in \S\ref{subsec:risk_anatomy}, the formal model sorts losses per flow to enable fine-grained risk control.
For each flow $f$, we sort its scenario losses $\{\ell_{f,q}\}_{q\in\mathcal{Q}}$ in descending order: $v_{f,r}$ denotes the $r$-th largest loss ($v_{f,1}\ge v_{f,2}\ge\cdots\ge v_{f,N}$), $I_{f,r}$ is a binary mask selecting which ranks contribute, and $\pi_{f,r}$ is the probability mass at rank $r$.
Crucially, we distinguish scenario index $q$ (identity) from rank index $r$ (position after sorting): if scenario $q$ lands at rank $r$, then $v_{f,r}=\ell_{f,q}$ and $\pi_{f,r}=p_q$.
The unified formulation over graph $\mathcal{G}(\mathcal{V}, \mathcal{E})$ with $N{=}|\mathcal{Q}|$ scenarios is:
\begin{subequations} \label{eq:unified_opt}
\begin{align}
\underset{\mathbf{x}, \boldsymbol{\delta}, \mathbf{v}, \boldsymbol{\pi}}{\text{minimize}} \quad
& \sum_{f \in \mathcal{F}} \sum_{r=1}^N \underbrace{\pi_{f,r} \cdot I_{f,r}}_{\text{Risk Mask}} \cdot v_{f,r} \label{eq:obj_unified} \\
\text{subject to} \quad
& \sum_{f} \sum_{t: e \in t} D_f \cdot x_{f,t} \le C_{e}, \quad \forall e \in \mathcal{E} \label{eq:const_capacity} \\
&x_{f,t}\ge0,\quad \forall f, t \\
& \ell_{f,q} = \max\left\{0, 1 - \sum_{t} x_{f,t} \cdot \alpha_{t,q}\right\}, \quad \forall f, q \label{eq:const_loss_def} \\
& \textbf{Sorting \& Permutation Constraints (Eq. \ref{eq:sort_logic})} \nonumber
\end{align}
\end{subequations}
\noindent Here $D_f$ is the demand size, $C_e$ is the capacity of link $e$, and $x_{f,t}$ defines the split ratios for flow $f$ across its tunnels $t\in\mathcal{T}_f$. Constraint \eqref{eq:const_capacity} imposes physical link capacity limits, while \eqref{eq:const_loss_def} maps a routing decision $\mathbf{x}$ to per-scenario loss $\ell_{f,q}$ via tunnel survival indicators $\alpha_{t,q}$.
The objective \eqref{eq:obj_unified} aggregates the \emph{sorted} loss vector via a rank-dependent mask.
To realize this, we introduce binary permutation variables $\delta_{f,q,r}\in\{0,1\}$ (equals 1 iff scenario $q$ is ranked $r$-th for flow $f$) and encode the sorting logic with standard Big-$M$ constraints:

\begin{subequations} \label{eq:sort_logic}
\begin{align}
&\sum_{r} \delta_{f,q,r} = 1, \quad \sum_{q} \delta_{f,q,r} = 1, \quad \forall f,q,r \label{eq:perm_matrix} \\
&v_{f,r} \le \ell_{f,q} + M(1-\delta_{f,q,r}), \quad \forall f,q,r \label{eq:bigM_upper} \\
&v_{f,r} \ge \ell_{f,q} - M(1-\delta_{f,q,r}), \quad \forall f,q,r \label{eq:bigM_lower} \\
&v_{f,r} \ge v_{f,r+1}, \quad \forall f,\, r=1,\ldots,N-1 \label{eq:sort_order} \\
&\pi_{f,r} = \sum_q p_q \cdot \delta_{f,q,r}, \quad \forall f,r \label{eq:perm_prob} \\
&\Gamma_{f,r} = \sum_{j=1}^r \pi_{f,j}, \quad \forall f,r \label{eq:cum_prob}
\end{align}
\end{subequations}
\noindent Constraint \eqref{eq:perm_matrix} makes $\boldsymbol{\delta}_f$ a permutation matrix that assigns each scenario $q$ to exactly one rank $r$ (and vice versa). Constraints \eqref{eq:bigM_upper}--\eqref{eq:bigM_lower} force $v_{f,r}$ to equal the loss $\ell_{f,q}$ of whichever scenario is assigned to rank $r$; \eqref{eq:sort_order} enforces that ranks are ordered by loss (largest to smallest). Here $M$ is a sufficiently large constant. Finally, \eqref{eq:perm_prob} and \eqref{eq:cum_prob} propagate scenario probabilities $p_q$ into ranked probability masses $\pi_{f,r}$ and their cumulative sum $\Gamma_{f,r}$.

Prior methods differ in both their rank selection policies and ranking granularity.
Scenario-level methods (FFC, TeaVaR) aggregate per-flow losses into a scenario loss $\ell_q$, then rank scenarios; all flows share a single mask ($I_{f,r} = I_r$).
Flow-level methods (FloMore) rank each flow independently; each has its own $I_{f,r}$.

\noindent \textbf{Case A: Robust} (FFC~\cite{ffc}). Scenario-level; dark mask. Only the worst scenario contributes:
\begin{equation}
    I_{f,1}=1, \quad I_{f,r}=0 \quad \forall f,\, r > 1.
\end{equation}
\noindent \textbf{Case B: Tail-Risk} (TeaVaR~\cite{teavar}). Scenario-level; blue mask. Selects the worst $1{-}\beta$ tail (CVaR):
\begin{equation}
    I_{f,r} = 
    \begin{cases} 
    1, & \text{if } \Gamma_{f,r} \le 1-\beta \\
    0, & \text{otherwise}
    \end{cases}
\end{equation}
\noindent \textbf{Case C: Quantile} (FloMore~\cite{flomore}). Flow-level. Each flow selects a single rank at the $\beta$-quantile (VaR):
\begin{equation}
    I_{f,r} = 1 \iff \Gamma_{f,r-1} < 1-\beta \le \Gamma_{f,r}.
\end{equation}
The amber mask in Figure~\ref{fig:risk_framework_analysis} illustrates a more general ``radical'' selection of the lower-loss tail; FloMore's quantile is a special case, selecting only the boundary rank.

\subsection{Motivation: From Combinatorial Search to Neural Approximation}
\label{sec:motivation_approx}

The unified formulation in \S\ref{sec:unified_formulation} reveals that solving risk-aware TE in this way (instance-by-instance optimization with explicit sorting constraints) is computationally prohibitive.
To overcome this, we propose a fundamental paradigm shift: \emph{Optimization as a Mapping.}
Fundamentally, any solver aims to realize an optimal mapping $\Psi^*$ that translates the network context (Topology $\mathcal{G}$, Demands $\mathbf{D}$, and Scenarios $\mathcal{Q}$) into an optimal allocation $\mathbf{x}^*$:
\begin{equation}
    \mathbf{x}^* = \Psi^*(\mathcal{G}, \mathbf{D}, \mathcal{Q}) = \arg\min_{\mathbf{x}} \mathcal{J}(\mathbf{x})
\end{equation}
\noindent where $\mathcal{J}(\mathbf{x})$ denotes the chosen risk objective (e.g., a tail-focused metric over scenario losses). Instead of computing this mapping implicitly by solving a combinatorial problem for every new input, we propose to \emph{approximate} it using a parameterized neural network $g_\theta$:
\begin{equation}
    \mathbf{x} = g_\theta(\mathcal{G}, \mathbf{D}, \mathcal{Q}) \approx \Psi^*(\mathcal{G}, \mathbf{D}, \mathcal{Q})
\end{equation}
where $\theta$ are learnable parameters. Intuitively, $g_\theta$ takes the network, demands, and scenario distribution as input and outputs a feasible routing decision without explicitly solving the problem online.
This shift from \textit{solving for variables} to \textit{learning a mapping} bypasses two barriers in the MILP formulation:
\begin{itemize}
    \item \textbf{Sorting.} The MILP optimizes over the permutation (which scenario gets which rank) via $O(|\mathcal{F}| \cdot N^2)$ binary variables $\boldsymbol{\delta}$—the main bottleneck. NeuroRisk does not: $g_\theta$ outputs $\mathbf{x}$ directly; ranking is only used when computing the loss (e.g., sort scenario losses then aggregate by the risk objective).
    \item \textbf{Linearization.} Tail-Risk and Quantile objectives (Mask $\times$ Loss) require Big-M linearization in MILP, adding many auxiliary variables. NeuroRisk optimizes the non-linear risk surface directly.
\end{itemize}

\section{Challenges of Deep Learning for Risk-Aware TE}
\label{sec:challenges}

Deep learning has achieved remarkable success in Traffic Engineering~\cite{dote,figret,harp,liu2025geminetlearningdualitybasediterative}: near-optimal routing with orders-of-magnitude speedup, generalizing across demands and topologies. These methods, however, are primarily designed for and benchmarked on Maximum Link Utilization (MLU), which is ill-suited for risk-aware settings: under failures, capacity becomes scarce and MLU provides no guidance on how to prioritize flows or distribute limited bandwidth. Risk-aware TE also often requires over-provisioning (split ratios $>$1) for failover, which the MLU formulation (split ratios sum to 1) does not model. Moreover, risk-aware TE reasons over a scenario set $\mathcal{Q}$ (which failures, with what probabilities), not just nominal demands and topology. Risk-aware objectives therefore require two capabilities that existing architectures lack: explicit capacity feasibility and scenario-level failure information. We elaborate on each next.

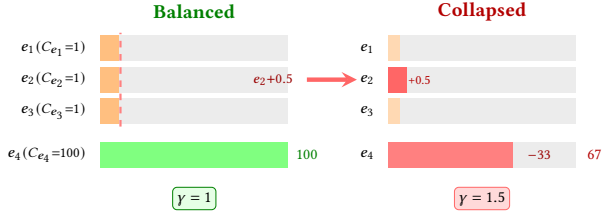
\begin{figure}[t]
\centering
\resizebox{0.95\columnwidth}{!}{%
\begin{tikzpicture}[>=stealth, font=\small]

\begin{scope}[local bounding box=left]
  \node[font=\small\bfseries, green!50!black] at (1.5, 3.4) {Balanced};
  
  \node[font=\scriptsize, anchor=east] at (-0.1, 2.8) {$e_1$\tiny$(C_{e_1}{=}1)$};
  \node[font=\scriptsize, anchor=east] at (-0.1, 2.3) {$e_2$\tiny$(C_{e_2}{=}1)$};
  \node[font=\scriptsize, anchor=east] at (-0.1, 1.8) {$e_3$\tiny$(C_{e_3}{=}1)$};
  \node[font=\scriptsize, anchor=east] at (-0.1, 1.1) {$e_4$\tiny$(C_{e_4}{=}100)$};
  
  \fill[orange!50] (0, 2.6) rectangle (0.3, 3.0);
  \fill[gray!15] (0.3, 2.6) rectangle (3, 3.0);
  \fill[orange!50] (0, 2.1) rectangle (0.3, 2.5);
  \fill[gray!15] (0.3, 2.1) rectangle (3, 2.5);
  \fill[orange!50] (0, 1.6) rectangle (0.3, 2.0);
  \fill[gray!15] (0.3, 1.6) rectangle (3, 2.0);
  
  \draw[red!50, dashed, line width=1pt] (0.32, 1.55) -- (0.32, 3.05);
  
  \fill[green!50] (0, 0.9) rectangle (3, 1.3);
  \node[font=\scriptsize, green!50!black] at (3.3, 1.1) {100};
  
  \node[font=\scriptsize, fill=green!10, draw=green!50!black, rounded corners=2pt, inner sep=2pt] at (1.5, 0.4) {$\gamma = 1$};
\end{scope}

\draw[->, line width=2pt, red!60] (3.3, 2.3) -- (4.1, 2.3);
\node[font=\tiny\bfseries, red!60!black, anchor=east] at (3.2, 2.3) {$e_2{+}0.5$};

\begin{scope}[shift={(4.6, 0)}, local bounding box=right]
  \node[font=\small\bfseries, red!70!black] at (1.5, 3.4) {Collapsed};
  
  \node[font=\scriptsize, anchor=east] at (-0.1, 2.8) {$e_1$};
  \node[font=\scriptsize, anchor=east] at (-0.1, 2.3) {$e_2$};
  \node[font=\scriptsize, anchor=east] at (-0.1, 1.8) {$e_3$};
  \node[font=\scriptsize, anchor=east] at (-0.1, 1.1) {$e_4$};
  
  \fill[orange!30] (0, 2.6) rectangle (0.2, 3.0);
  \fill[gray!15] (0.2, 2.6) rectangle (3, 3.0);
  \fill[orange!30] (0, 1.6) rectangle (0.2, 2.0);
  \fill[gray!15] (0.2, 1.6) rectangle (3, 2.0);
  
  \fill[red!60] (0, 2.1) rectangle (0.3, 2.5);
  \fill[gray!15] (0.3, 2.1) rectangle (3, 2.5);
  \node[font=\tiny, red!70!black] at (0.5, 2.3) {+0.5};
  
  \fill[red!50] (0, 0.9) rectangle (2, 1.3);
  \fill[gray!15] (2, 0.9) rectangle (3, 1.3);
  \node[font=\scriptsize, red!70!black] at (3.3, 1.1) {67};
  
  \node[font=\tiny, red!60!black, anchor=west] at (2.1, 1.1) {$-33$};
  
  \node[font=\scriptsize, fill=red!15, draw=red!60, rounded corners=2pt, inner sep=2pt] at (1.5, 0.4) {$\gamma = 1.5$};
\end{scope}

\end{tikzpicture}%
}
\caption{\textbf{GS optimization fragility.} Consider bottleneck edges $e_1$--$e_3$ (capacity $C_e{=}1$) and a large edge $e_4$ ($C_{e_4}{=}100$). The optimizer must keep all $e_i$ loads at their capacity limits. When $e_2$'s load exceeds by just 0.5, $\gamma$ rises to 1.5 and \emph{all} edges are scaled--$e_4$ loses 33 units to accommodate $e_2$'s 0.5 excess. This disproportionate collateral damage makes convergence difficult.}
\label{fig:gs_example}
\end{figure}

\noindent\textbf{The Feasibility Barrier (output).}
Existing methods use \emph{global scaling} (GS) to satisfy constraints \eqref{eq:const_capacity}: divide all allocations by $\gamma_{\max} = \max_e (\sum_{f} \sum_{t: e \in t} D_f \cdot x_{f,t} / C_e)$. GS creates a ``whack-a-mole'' pathology (Figure~\ref{fig:gs_example}): the model must keep \emph{every} bottleneck edge at or below capacity; overshooting any one (e.g., $e_2$ by 0.5) collapses throughput network-wide ($e_4$ loses 33 units).
Furthermore, GS obscures the true `offending' edges by normalizing them to feasibility, which inadvertently encourages the model to increase their allocations. This triggers a cycle of shifting bottlenecks and throughput oscillations, fundamentally stalling convergence.

\noindent\textbf{The Representation Barrier (input).}
Risk-aware TE requires scenario-level failure information: which failures, with what probabilities. Existing Deep TE methods accept demands and topology but lack a mechanism to ingest a scenario set $\mathcal{Q}$. To our knowledge, FauTE~\cite{faute} is the only attempt, encoding per-link failure probabilities $p_e$ as a soft penalty $\sum_e p_e \cdot \sum_{f} \sum_{t: e \in t} D_f \cdot x_{f,t}$. Collapsing scenarios into scalar probabilities loses correlation structure (e.g., links sharing conduits often fail together) and restricts optimization to expectations. Existing risk objectives (e.g., CVaR) all require reasoning over the loss distribution across discrete scenarios.

These two barriers are fundamental mismatches between existing Deep TE architectures and risk-aware optimization. Overcoming them necessitates a holistic redesign of both the output space (for feasibility) and the input representation (for scenario encoding).

\section{System Design}
\label{sec:design}

\subsection{Overview}
\label{subsec:overview}

\begin{figure*}[t]
    \centering
    \includegraphics[width=\textwidth]{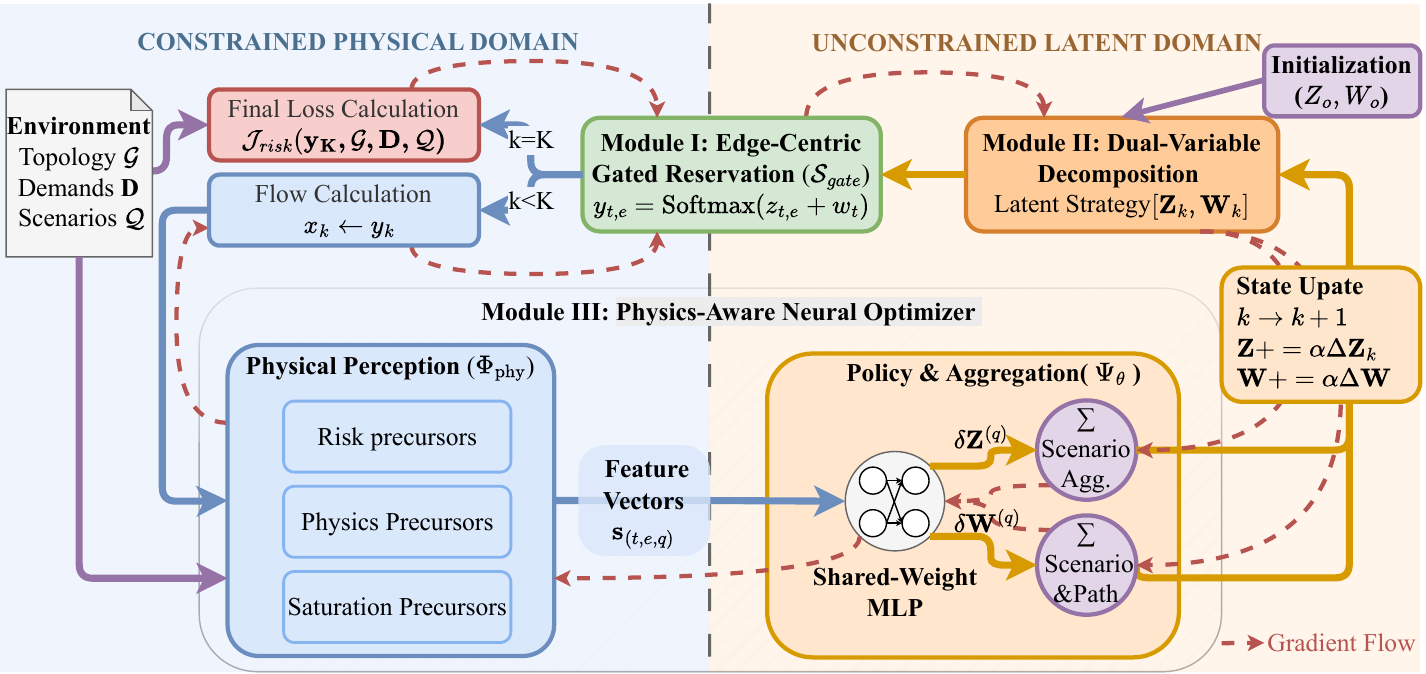}
    \caption{\textbf{NeuroRisk framework architecture.}
    The system separates into two domains: the \textbf{Unconstrained Latent Domain} (right, orange) where optimization occurs freely, and the \textbf{Constrained Physical Domain} (left, blue) where feasibility is enforced.
    \textbf{Module I} maps latent variables $(\mathbf{Z}, \mathbf{W})$ to strictly feasible reservations $\mathbf{y}$ via per-edge Softmax.
    \textbf{Module II} maintains two coupled latent variables: local contention $\mathbf{Z}$ and tunnel-level gating $\mathbf{W}$.
    \textbf{Module III} extracts gradient-proxy features from $(\mathbf{y}, \mathcal{Q})$ and computes updates via a shared-weight MLP, aggregated across scenarios.}
    \label{fig:framework}
\end{figure*}

To dismantle these barriers, NeuroRisk fundamentally rethinks the TE decision space and the optimization paradigm.
To address the Feasibility Barrier, we shift the decision variables from globally-coupled \textbf{tunnel-level allocations $\mathbf{x}$} to \textbf{edge-local reservation ratios $\mathbf{y}$}, which guaranties capacity feasibility by design without post-hoc scaling.

To overcome the Representation Barrier, we exploit a key property of \emph{Sort-and-Select} objectives: the total gradient decomposes additively across scenarios, i.e., $\nabla_{\mathbf{y}} \mathcal{J} = \sum_q \rho_q \nabla_{\mathbf{y}} \mathcal{J}_q$, where $\rho_q = p_q \cdot m_q$ combines the scenario probability $p_q$ with the risk-specific selection mask $m_q$ (e.g., $m_q = 1$ for tail scenarios in CVaR, $0$ otherwise), and $\mathcal{J}_q$ is the risk contribution from scenario $q$. Crucially, this property holds for the gradient \emph{with respect to the decision variables} $\mathbf{y}$, not the neural parameters $\theta$. This distinction dictates a fundamental shift from \emph{one-shot prediction} to \emph{iterative optimization}:

\begin{itemize}
    \item \textbf{The ``Blank Slate'' Dilemma (One-shot):} A direct predictor lacks an intermediate decision state. Without a concrete $\mathbf{y}$, the gradient $\nabla_{\mathbf{y}} \mathcal{J}_q$ is undefined. Consequently, the network must implicitly encode the combinatorial interactions of all scenarios into its weights--a massive representation bottleneck.
    \item \textbf{The ``State Sensitivity'' Advantage (Iterative):} NeuroRisk maintains an explicit decision anchor $\mathbf{y}_k$ at each step. This allows us to compute the \emph{state sensitivity}---how the risk in each scenario $q$ responds to changes in the current plan. These per-scenario sensitivities are \textbf{independent and additive}: scenario $q_1$ pushes $\mathbf{y}_k$ one way; scenario $q_2$ pushes it another. NeuroRisk simply aggregates these local signals, decoupling the architecture from the complexity of $\mathcal{Q}$.
\end{itemize}

Leveraging there insights, NeuroRisk is formulated as a learnable dynamical system~\cite{liu2025geminetlearningdualitybasediterative,10.5555/3104322.3104374,NIPS2016_1679091c,RAISSI2019686} (Figure~\ref{fig:framework}) that evolves a latent strategy $(\mathbf{Z}, \mathbf{W})$ over $K$ iterations. Each step $k$ executes a synchronized Generate-Perceive-Update cycle:
\begin{enumerate}
    \item \textbf{Generate:} A gated projection $\mathcal{S}_{\text{gate}}$ maps the latent state $(\mathbf{Z}_k, \mathbf{W}_k)$ to a physically feasible reservation $\mathbf{y}_k$;
    \item \textbf{Perceive:} A physics engine $\Phi_{\text{phy}}$ computes current allocations and extracts gradient-aligned features $\mathbf{s}_{k,q}$ for each scenario $q \in \mathcal{Q}$;
    \item \textbf{Update:} A shared neural policy $\Psi_\theta$ processes these features to propose local updates, which are aggregated using risk weights $\rho_q$ to refine the global strategy.
\end{enumerate}
The resulting computation graph is fully differentiable, enabling end-to-end training where the policy $\theta$ learns to minimize the terminal risk $\mathcal{J}$ by backpropagating:
\begin{subequations}
\label{eq:neurorisk_dynamics}
\begin{align}
\min_{\theta} \quad & \mathcal{J}( \mathbf{y}_K; \mathcal{G}, \mathbf{D}, \mathcal{Q} ) \label{eq:objective}\\
\text{s.t.} \quad & \mathbf{y}_k = \mathcal{S}_{\text{gate}}(\mathbf{Z}_k, \mathbf{W}_k) \label{eq:generate}\\
& (\Delta \mathbf{Z}_k, \Delta \mathbf{W}_k) = \sum_{q \in \mathcal{Q}} \rho_q \cdot \Psi_{\theta} \big( \Phi_{\text{phy}}( \mathbf{y}_k, \mathcal{G}, \mathbf{D}, q ) \big) \label{eq:update}\\
& \mathbf{Z}_{k+1} = \mathbf{Z}_k + \Delta \mathbf{Z}_k, \; \mathbf{W}_{k+1} = \mathbf{W}_k + \Delta \mathbf{W}_k \label{eq:evolve}
\end{align}
\end{subequations}
Notably, $\Psi_\theta$ operates exclusively on physics-aware cues $\mathbf{s}_{k,q}$ and never observes raw topology or scenario data. By learning the underlying update dynamics rather than memorizing instance-specific mappings, NeuroRisk achieves \emph{zero-shot generalization} to unseen network scales and failure distributions.
The following sections detail each component: \S\ref{subsec:gated_reservation} introduces Gated Reservation ($\mathcal{S}_{\text{gate}}$), addressing the Feasibility Barrier; while \S\ref{subsec:perception} describes the Physics Engine ($\Phi_{\text{phy}}$), addressing the Representation Barrier. 

\subsection{Gated Reservation: Decision Space Transformation}
\label{subsec:gated_reservation}

As analyzed in \S\ref{sec:challenges}, Global Scaling (GS) suffers from optimization fragility: a single bottleneck's micro-fluctuation triggers network-wide collapse. A natural fix is Local Scaling (LS): instead of a shared denominator $\gamma_{\max}$, each tunnel $t$ is scaled by its own bottleneck factor $\gamma_t = \max_{e \in t}(\sum_{f} \sum_{\tau: e \in \tau} D_f \cdot x_{f,\tau} / C_e)$. This localizes the impact--adjusting one tunnel no longer directly affects others. Unfortunately, LS introduces a subtler but equally fatal problem: \emph{gradient distortion}.

\begin{figure}[t]
\centering
\resizebox{0.65\columnwidth}{!}{%
\begin{tikzpicture}[scale=1.2, >=Stealth]

    \draw[->] (0,0) -- (5.5,0);
    \draw[->] (0,0) -- (0,4.2);
    
    \coordinate (Opt) at (1.8, 1.8);
    
    \coordinate (E1_End) at (1.3, 4.0);
    
    \coordinate (E2_End) at (5.5, 2.15);
    
    \coordinate (Ridge_End) at (5.0, 4.2);
    
    \fill[gray!20] (0,0) -- (5.5, 0) -- (E2_End) -- (Opt) -- (E1_End) -- (0, 4.0) -- cycle;
    \node[font=\footnotesize, gray!70!black, fill=white, fill opacity=0.85, text opacity=1, inner sep=1.2pt] at (0.95,0.55) {Feasible region};
    \node[font=\footnotesize, gray!70!black, fill=white, fill opacity=0.85, text opacity=1, inner sep=1.2pt] at (4.35,3.05) {Infeasible region};
    
    \draw[blue!70, line width=2pt] (Opt) -- (E1_End);
    \node[font=\footnotesize, blue!70, left] at (1.05, 3.5) {$e_1$};
    \draw[red!70, line width=2pt] (Opt) -- (E2_End);
    \node[font=\footnotesize, red!70, below] at (4.5, 2.0) {$e_2$};
    
    \draw[gray, dashed, line width=1.2pt] (Opt) -- (Ridge_End);
    \node[font=\footnotesize, gray, fill=white, fill opacity=0.9, text opacity=1, inner sep=1.2pt] at (4.55, 3.78) {Ridge};
    
    \fill[green!50!black] (Opt) circle (4pt);
    \node[font=\small, green!50!black, below] at ($(Opt)+(0,-0.15)$) {$x^*$};
    
    \coordinate (A) at (3.0, 2.95);
    
    \coordinate (ProjA) at ($(Opt)!(A)!(E1_End)$);
    
    \fill[blue!70] (A) circle (3pt);
    \node[font=\small, blue!70, above] at ($(A)+(0,0.1)$) {$A$};
    \draw[blue!70, dashed, line width=1pt] (A) -- (ProjA);
    \fill[blue!70] (ProjA) circle (3pt);
    \node[font=\small, blue!70, left] at ($(ProjA)+(-0.1,0)$) {$A'$};
    \draw pic[draw=blue!70, angle radius=5pt, line width=0.8pt] {right angle = A--ProjA--E1_End};
    
    \coordinate (B) at (3.3, 2.65);
    
    \coordinate (ProjB) at ($(Opt)!(B)!(E2_End)$);
    
    \fill[red!70] (B) circle (3pt);
    \node[font=\small, red!70, above right] at ($(B)+(0.05,0.05)$) {$B$};
    \draw[red!70, dashed, line width=1pt] (B) -- (ProjB);
    \fill[red!70] (ProjB) circle (3pt);
    \node[font=\small, red!70, below] at ($(ProjB)+(0,-0.15)$) {$B'$};
    \draw pic[draw=red!70, angle radius=5pt, line width=0.8pt] {right angle = B--ProjB--E2_End};
    
    \draw[orange, ->, line width=1.5pt] (A) -- (B);
    
    \draw[blue!70, <->, line width=0.8pt] (Opt) -- (ProjA);
    \node[font=\scriptsize, blue!70, fill=white, inner sep=0.5pt] at ($(Opt)!0.5!(ProjA)+(0.15,0.15)$) {$d_A$};
    
    \draw[purple, <->, line width=0.8pt] (Opt) -- (ProjB);
    \node[font=\scriptsize, purple, fill=white, inner sep=0.5pt] at ($(Opt)!0.5!(ProjB)+(0,-0.2)$) {$d_B > d_A$};

\end{tikzpicture}%
}
\caption{\textbf{Mechanism of Projection Discontinuity.}}
\label{fig:ls_projection}
\end{figure}
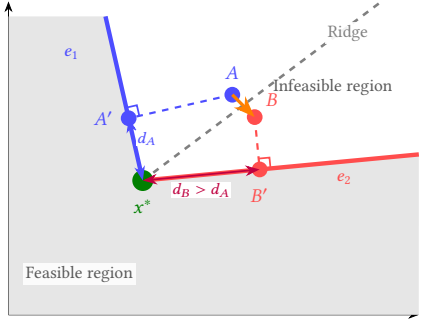
LS enforces feasibility via a projection operator $\pi(\cdot)$ that maps an infeasible solution back onto the constraint boundary. The key issue is that each tunnel variable $x_{f,t}$ appears in multiple capacity constraints (one per edge along its path). As the optimizer explores, different constraints become active depending on the current solution, and the effective projection rule switches accordingly. Figure~\ref{fig:ls_projection} illustrates this Projection Discontinuity Trap: a small step that appears beneficial in the unconstrained domain ($A{\to}B$) can cross the Ridge and trigger a jump in the projection ($A'=\pi(A)$ to $B'=\pi(B)$), leading to worse outcomes and a bounce-back that prevents progress toward $x^*$. 

In our experiments, LS-based training exhibits a strong short-path preference: when both short and long paths are feasible, LS tends to concentrate traffic on shorter (often direct) routes, since longer paths traverse more edges and are more likely to activate tight constraints and incur projection penalties. For risk-aware TE, this bias is catastrophic: backup paths are essential for absorbing failures, yet LS systematically under-utilizes them.

Both GS and LS fail because they attempt to enforce feasibility via post-hoc scaling. The root cause is that flow-level variables are globally coupled: each $x_t$ appears in multiple constraints, so any scaling operation must navigate this entanglement. The solution is not a better scaling rule--it is to \emph{transform the decision space} so that feasibility is guaranteed by construction, without any scaling.

\subsubsection{Module I: Edge-Centric Reservation}
\label{subsec:module1}

We implement this decision-space transformation via Bandwidth Reservation (BR): by directly optimizing per-edge reservation ratios, feasibility is guaranteed by construction, eliminating the need for post-hoc scaling.
Traditional TE asks: ``What fraction of flow $f$'s demand goes on tunnel $t$?'' This produces variables $x_{f,t}$ that must satisfy capacity constraints $\sum_{f,t: e \in t} D_f \cdot x_{f,t} \le C_e$ for every edge. The problem: each $x_{f,t}$ appears in \emph{multiple} constraints (one per edge along the tunnel), so adjusting one variable ripples through many constraints. We invert the perspective: ``What fraction of edge $e$'s capacity is reserved for tunnel $t$?'' This produces variables $y_{t,e} \in [0,1]$ with constraint $\sum_{t \in \mathcal{T}_e} y_{t,e} = 1$ for each edge, where $\mathcal{T}_e = \{t : e \in t\}$ denotes the set of tunnels traversing edge $e$. Now each $y_{t,e}$ appears in \emph{exactly one} constraint--its own edge's partition. Edges allocate capacity independently; there is no global coupling.
A natural implementation is per-edge Softmax over learnable logits:
\begin{equation}
\label{eq:naive_softmax}
\tilde{y}_{t,e} = \frac{\exp(z_{t,e})}{\sum_{\tau \in \mathcal{T}_e} \exp(z_{\tau,e})}, \quad \forall t \in \mathcal{T}_e,
\end{equation}
where $z_{t,e}$ is a learnable logit for tunnel $t$ on edge $e$. By construction, $\tilde{y}_{t,e} \ge 0$ and $\sum_t \tilde{y}_{t,e} = 1$---capacity is never violated, no scaling needed. This eliminates both GS's collateral damage (no shared $\gamma_{\max}$) and LS's projection trap (no per-tunnel scaling factor). The bandwidth allocation ratio can be recovered as: 
\begin{equation}
    x_{f,t}= (\min_{e \in t}C_e \cdot \tilde{y}_{t,e})/D_f
\end{equation}

However, this naive approach has a critical flaw: the \emph{Short-Board Trap}. A tunnel's delivered bandwidth is $\min_{e \in t} C_e \cdot \tilde{y}_{t,e}$---the minimum across all edges. If the optimizer increases $\tilde{y}_{t,e_1}$ on one edge while others remain low, the extra allocation is ``wasted'' (throughput unchanged since the bottleneck is elsewhere). Competing tunnels on $e_1$ then push to reclaim the unused capacity. Path-level growth---raising all edges of a tunnel together---is prohibitively difficult with purely edge-local $z_{t,e}$ updates. Each edge's Softmax is an independent competition; there is no mechanism for ``raise my share on \emph{all} my edges simultaneously.''
To escape this trap, we need to restore tunnel-level coordination---but without reintroducing the projection discontinuities of LS.

\subsubsection{Module II: Tunnel-Level Gating}
\label{subsec:module2}

To escape the Short-Board Trap, we introduce a tunnel-level gate $w_t$ that coordinates all edges of a tunnel. Combined with the edge-level Softmax from Module I, this defines the gated projection $\mathcal{S}_{\text{gate}}$ formalized in the overview. The final reservation ratio $y_{t,e}$ is computed as:
\begin{equation}
\label{eq:gated_softmax}
y_{t,e} = \frac{\exp(z_{t,e} + w_t)}{\sum_{\tau \in \mathcal{T}_e} \exp(z_{\tau,e} + w_{\tau})}, \quad \forall t \in \mathcal{T}_e.
\end{equation}
The two latent variables serve complementary roles: \emph{Local contention} ($z_{t,e}$) controls how aggressively tunnel $t$ competes on edge $e$, relative to other tunnels sharing that edge; \emph{Global gating} ($w_t$) is a tunnel-wide priority score shared across all edges of $t$. Because $w_t$ appears in the Softmax on \emph{every} edge along tunnel $t$, increasing $w_t$ simultaneously boosts $t$'s share on all its edges---exactly the ``raise all boards together'' signal that pure $z_{t,e}$ cannot express.

Think of a tunnel as a barrel: its throughput is limited by the \emph{shortest board} (bottleneck edge). With edge-local $z_{t,e}$ alone, raising one board while others remain short wastes capacity---the water still spills. The tunnel-level gate $w_t$ acts as a \emph{lifting mechanism} that raises all boards simultaneously. Mathematically, the gradient $\partial \mathcal{J}/\partial w_t$ aggregates feedback from \emph{all} edges along tunnel $t$:
\begin{equation}
\label{eq:w_gradient}
\frac{\partial \mathcal{J}}{\partial w_t} = \sum_{e \in t} \frac{\partial \mathcal{J}}{\partial y_{t,e}} \cdot \frac{\partial y_{t,e}}{\partial w_t}
\end{equation}
Crucially, the summation $\sum_{e \in t}$ in Eq.~\eqref{eq:w_gradient} acts as a voting mechanism: globally efficient tunnels accumulate positive votes across all their edges and grow uniformly; inefficient detours accumulate negative votes and shrink.
In LS, the decision variables are already tunnel-level (flow/tunnel bandwidth allocations). Feasibility is enforced by a \emph{per-tunnel scaling/projection factor} $\gamma_t$, which is \emph{computed} from edge loads via a non-smooth $\max$ over constraints. This creates projection discontinuities. In contrast, $w_t$ is a \emph{learnable} tunnel-level coordination variable that lives inside the smooth Softmax, ensuring stable gradient propagation.

\noindent\textbf{Summary: The Best of Both Worlds.}
Gated Reservation synthesizes the strengths of BR and LS. It inherits the strictly feasible, smooth optimization landscape of the edge-centric design (Module I), while $w_t$ restores the vital \emph{tunnel-level coordination} of LS. This synergy avoids the optimization traps of post-hoc scaling, effectively enabling the ``raise all boards together'' capability---a collective growth mechanism that neither approach can achieve in isolation.

\subsection{Physics-Aware Neural Optimizer (Module III)}
\label{subsec:perception}

\begin{table}[t]
\centering
\small
\renewcommand{\arraystretch}{1.1}
\begin{tabular}{@{}llp{0.6\columnwidth}@{}}
\toprule
Category & Feature & Physical Interpretation \\
\midrule
\multirow{3}{*}{\textit{Risk}} 
  & $\alpha_{t,q}$ & Tunnel survival status (1 if path connects, 0 if broken) \\
  & $\ell_{f,q}$ & Flow $f$'s unserved demand ratio under $q$ \\
  & $\ell_q$ & Scenario severity (global loss) \\
\midrule
\multirow{4}{*}{\textit{Physics}} 
  & $x_{f,t}$ & Split ratios of tunnel $t$ for flow $f$ \\
  & $m_{t,e}$ & \emph{Bottleneck Margin}: $C_e y_{t,e} - b_t$ \\
  & $C_e$ & Edge capacity (normalized) \\
  & $D_{f(t)}$ &  Demand magnitude\\
\midrule
\textit{Saturation} & $y_{t,e}$ & Current Softmax share (gradient saturation proxy) \\
\bottomrule
\end{tabular}
\caption{Physics-Aware Feature Vector $\mathbf{s}_{t,e,q}$. These features directly correspond to the components in Eq.~\eqref{eq:micro_gradient}.}
\label{tab:features}
\end{table}
Module III functions as the \emph{neural update policy} within the iterative optimization loop. As illustrated in Figure~\ref{fig:framework}, it comprises two coordinated sub-components: (1) a \textbf{Physics Engine ($\Phi_{\text{phy}}$)} that extracts gradient-relevant features from the physical state, and (2) a \textbf{Shared-Weight Policy ($\Psi_\theta$)} that maps these features to optimal update steps. At each iteration $k$, this module observes the feasible reservation $\mathbf{y}_k$ and failure scenarios $\mathcal{Q}$, then outputs incremental updates $(\Delta \mathbf{Z}_k, \Delta \mathbf{W}_k)$ to evolve the latent strategy.

\noindent\textbf{Physics Engine: Gradient-Grounded Feature Design.}
As established in \S\ref{subsec:overview}, maintaining an explicit state $\mathbf{y}_k$ allows us to compute the \emph{state sensitivity}---how each scenario's risk responds to changes in the latent variables $(\mathbf{Z}, \mathbf{W})$. We now detail how to translate this analytical structure into learnable features.
Rather than learning from raw data, we ground our per-scenario features $\mathbf{s}_{t,e,q}$ in the analytical gradient, which captures exactly the information needed to improve the objective. By the chain rule, the gradient with respect to the edge-level latent variable $z_{t,e}$ factorizes into three physically interpretable components (Risk $\times$ Physics $\times$ Saturation):
\begin{equation}
\label{eq:micro_gradient}
    \frac{\partial \mathcal{J}_q}{\partial z_{t,e}} =
    \underbrace{\frac{\partial \mathcal{J}_q}{\partial x_{f,t}}}_{\text{Risk}}
    \cdot
    \underbrace{\frac{\partial x_{f,t}}{\partial y_{t,e}}}_{\text{Physics}}
    \cdot
    \underbrace{\frac{\partial y_{t,e}}{\partial z_{t,e}}}_{\text{Saturation}}
\end{equation}
The gradient with respect to the tunnel-level gate $w_t$ has the same structure, summed over all edges along the tunnel ($\sum_{e \in t}$).
(1) \emph{Risk Sensitivity} captures how loss responds to allocation changes; (2) \emph{Physical Coupling} ($\partial x/\partial y$) captures the ``Short-Board'' logic---it is non-zero only if edge $e$ is the bottleneck, motivating the \emph{Bottleneck Margin} feature $m_{t,e}$; (3) \emph{Saturation State} captures the Softmax derivative.
The Physics Engine $\Phi_{\text{phy}}$ uses these components not to regress the gradient directly, but to construct a physics-aware state representation (summarized in Table~\ref{tab:features}; derivation in Appendix~\ref{app:physics_features}) that enables the policy to learn update directions.

\noindent\textbf{Update Dynamics.}
The neural policy $\Psi_\theta$ is a lightweight MLP applied to each feature tuple $\mathbf{s}_{t,e,q}$. Per-scenario outputs are aggregated using the risk weights $\rho_q$ defined in \S\ref{subsec:overview}:
\begin{subequations}
\label{eq:update_dynamics}
\begin{align}
\Delta z_{t,e} &= \sum_{q \in \mathcal{Q}} \rho_q \cdot \Psi_\theta^{(z)}(\mathbf{s}_{t,e,q}) \label{eq:update_z} \\
\Delta w_t &= \sum_{e \in t} \sum_{q \in \mathcal{Q}} \rho_q \cdot \Psi_\theta^{(w)}(\mathbf{s}_{t,e,q}) \label{eq:update_w}
\end{align}
\end{subequations}
Eq.~\eqref{eq:update_z} updates local contention using edge-specific signals, while Eq.~\eqref{eq:update_w} sums across the tunnel ($\sum_{e \in t}$) to adjust the tunnel-wide gate. By operating on these physics-derived features rather than raw topology, the learned policy $\Psi_\theta$ captures the invariant \emph{update dynamics} of the optimization process, enabling zero-shot generalization to unseen topologies and scenario distributions.

\section{Evaluation}
\label{sec:evaluation}

We outline methodology (\S\ref{subsec:methodology}), compare the quality and runtime of the solution (\S\ref{subsec:eval_accuracy} -- \S\ref{subsec:eval_runtime}), evaluate generalization across scenarios, demands, tunnels and topologies (\S\ref{subsec:eval_generalization}), and provide ablation analysis (\S\ref{subsec:ablation}).

\subsection{Methodology}
\label{subsec:methodology}

\begin{figure*}[tbh]
\centering
\begin{minipage}{0.49\textwidth}
    \centering
    \includegraphics[width=\textwidth]{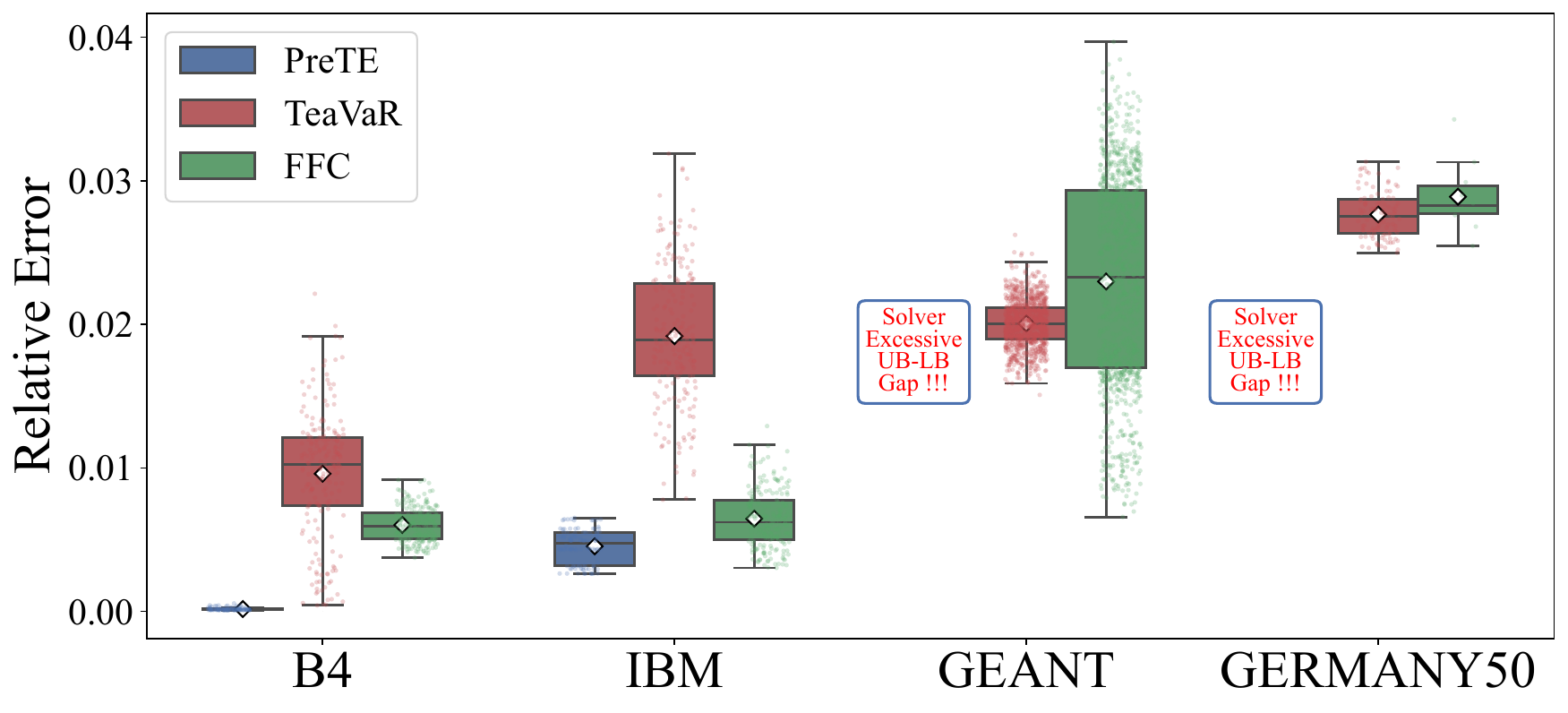}
    \vspace{-0.75em}
    {\small\textbf{(a)} Relative error vs.\ Gurobi optimum.}
\end{minipage}\hfill
\begin{minipage}{0.49\textwidth}
    \centering
    \includegraphics[width=\textwidth]{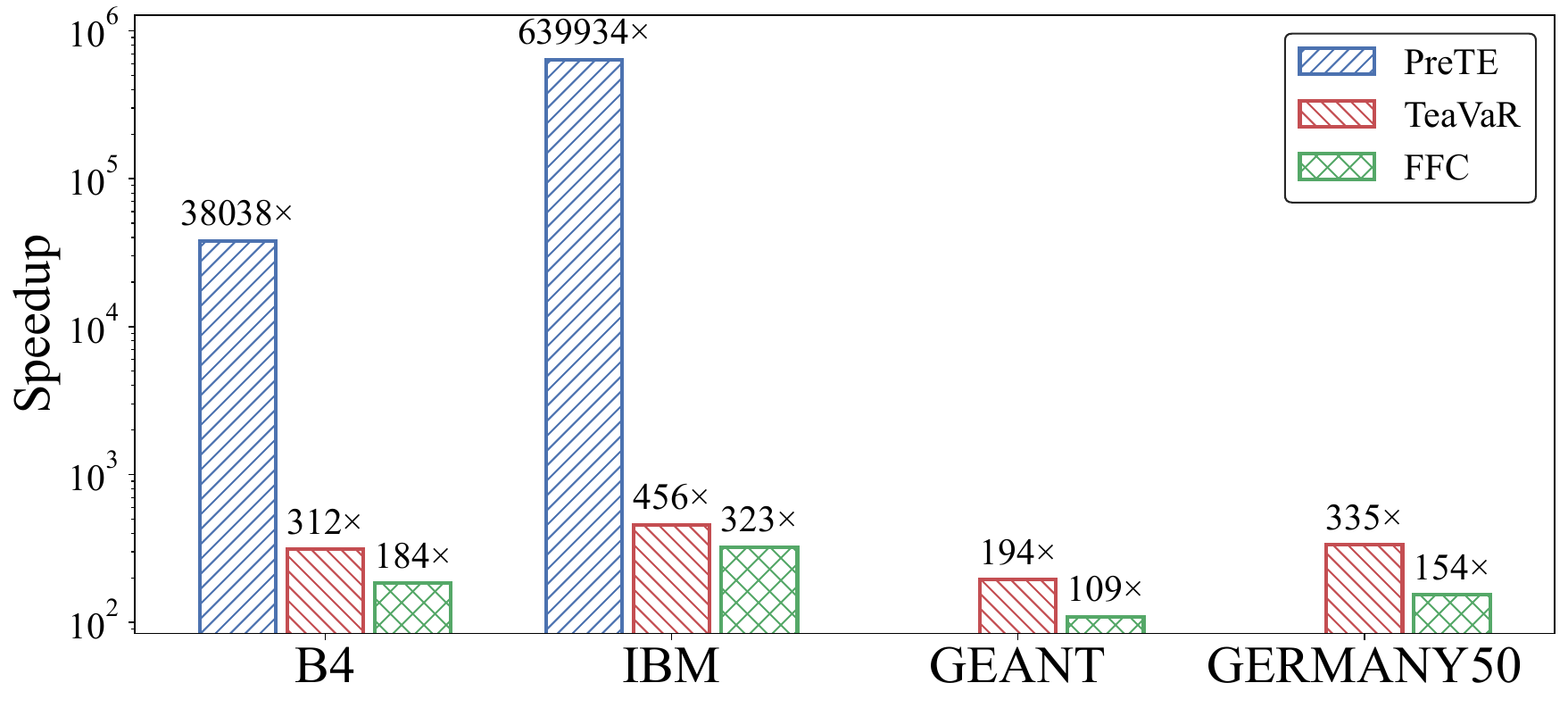}
    \vspace{-0.75em}
    {\small\textbf{(b)} End-to-end speedup (log-scale).}
\end{minipage}
\caption{\textbf{Overall performance summary.} (a) Relative error of NeuroRisk versus the Gurobi optimum for each objective (PreTE/TeaVaR/FFC). \emph{PreTE on \textsc{GEANT} and \textsc{GERMANY50} is omitted (N/A)} because the solver fails to obtain a non-trivial solution within a 1-hour time budget (unmet ratio remains 1.0). (b) End-to-end speedup of NeuroRisk inference over solver-based baselines, averaged per topology and objective (including model construction time). \textbf{For \textsc{GERMANY50}, GPU latency is measured with scenario chunking to cap peak memory.}}
\label{fig:overall_summary}
\end{figure*}

\noindent\textbf{Topologies.}
Our evaluation covers five WAN topologies representing diverse scales and connectivity patterns: \textsc{B4}, \textsc{IBM}, \textsc{GEANT}, \textsc{GERMANY50}, and \textsc{TATANID}.
For \textsc{B4} and \textsc{IBM}, we use the well-known optical-layer topologies also used in Semi-Oblivious TE~\cite{Semi-Oblivious} and generate IP-layer topologies using the distributions described in ARROW~\cite{ARROW}.
For \textsc{GEANT}, \textsc{GERMANY50} and \textsc{TATANID}, we source optical-layer topologies from the TopologyBench Dataset~\cite{matzner2024topology} and derive IP-layer topologies via a rule-based optical-to-IP mapping.
Unless otherwise stated, candidate tunnels are generated by $K_{sp}$-shortest paths (KSP) with $K_{sp}{=}3$ per source-destination pair.
Table~\ref{tab:topology} summarizes the information for each topology.
\begin{table}[tbh]
\centering
\small
\begin{tabular}{@{}lccc@{}}
\toprule
Topology & \#Nodes & \#Edges & \#Tunnels \\
\midrule
\textsc{B4} & 12 & 55 & 396 \\
\textsc{IBM} & 17 & 85 & 816 \\
\textsc{GEANT} & 22 & 36 & 1,386 \\
\textsc{GERMANY50} & 50 & 88 & 7,350 \\
\textsc{TATANID} & 142 & 180 & 60,066 \\
\bottomrule
\end{tabular}
\caption{Network topologies in our evaluation.}
\label{tab:topology}
\end{table}

\noindent\textbf{Workloads, scenarios, and setup.}
For \textsc{B4} and \textsc{IBM}, we use traffic matrices from PreTE~\cite{prete}; for \textsc{GEANT}, \textsc{GERMANY50}, and \textsc{TATANID}, we generate traffic matrices using a standard gravity model~\cite{applegateMakingIntraDomainRouting,roughanExperienceMeasuringInternet2002}.
We follow PreTE~\cite{prete} to generate a probabilistic multi-link scenario set $\mathcal{Q}$ with pruning cutoff $\tau$, and parameterize each scenario configuration by a pair $(c,s)$, where the cutoff is set to $\tau\triangleq c \times  10^{-5}$ and $\lambda \triangleq s \times 10^{-3} $ is the Weibull scale.
We train NeuroRisk on $(100,2)$, $(50,2)$, and $(50,4)$, and reserve all other $(c,s)$ combinations for testing.

We compare against \textbf{FFC}~\cite{ffc}, \textbf{PreTE}~\cite{prete}, \textbf{TeaVaR}~\cite{teavar}, and \textbf{FauTE}~\cite{faute}; solver-based baselines are solved using Gurobi 9.5.1~\cite{gurobi} and we report TotalTime.
Unless otherwise stated, we set $\beta{=}0.95$ (quantile for PreTE, CVaR level for TeaVaR).
Experiments run on two servers: Gurobi optimization on an Intel\textsuperscript{\textregistered} Xeon\textsuperscript{\textregistered} Gold 6530 CPU (1.5~TB RAM); DL Models inference on an Intel\textsuperscript{\textregistered} Xeon\textsuperscript{\textregistered} Platinum 8558 CPU (2.0~TB RAM) with an NVIDIA H200 GPU (141~GB VRAM). NeuroRisk is implemented in PyTorch 2.4.1 (CUDA 12.1) with Python 3.8.0; key hyperparameters are in Appendix~\ref{app:hyperparams}.

\subsection{Overall Performance}
\label{subsec:eval_accuracy}

We summarize end-to-end solution quality and speed across topologies and risk objectives, and compare against FauTE, a neural baseline that collapses the scenario set.

\noindent\textbf{Takeaway.}
Figure~\ref{fig:overall_summary}(a) shows that NeuroRisk achieves small optimality gaps across objectives and topologies where solver baselines are tractable, while Figure~\ref{fig:overall_summary}(b) shows orders-of-magnitude end-to-end speedups that make frequent reoptimization practical.
For flow-level objectives (PreTE) on larger topologies (\textsc{GEANT}, \textsc{GERMANY50}), the solver times out (1 hour) with a trivial solution (zero throughput); NeuroRisk produces a non-trivial feasible routing in ${\sim}$18–68 ms.


\noindent\textbf{Why scenario collapse is insufficient (FauTE).}
\label{subsec:eval_faute}
We compare against FauTE~\cite{faute}, a neural baseline that collapses $\mathcal{Q}$ into link-local failure probabilities and optimizes a surrogate objective; for a fair comparison, we evaluate both methods on expected throughput over $\mathcal{Q}$ and use the LP optimum from Gurobi as ground truth.
Table~\ref{tab:faute_throughput} summarizes results on \textsc{B4} and \textsc{IBM}, revealing substantial performance gaps for the surrogate-based approach.

\begin{table}[tbh]
\centering
\small
\begin{tabular}{@{}lccc@{}}
\toprule
Topology & Method & Expected Throughput & Rel.\ errors \\
\midrule
\textsc{B4} & Solver (Gurobi) & 73.21k & 0 \\
& NeuroRisk (Ours) & 72.94k & 0.28\% \\
& FauTE~\cite{faute} & 48.73k & 33.37\% \\
\midrule
\textsc{IBM} & Solver (Gurobi) & 94.54k & 0 \\
& NeuroRisk (Ours) & 93.40k & 1.23\% \\
& FauTE~\cite{faute} & 27.71k & 70.50\% \\
\bottomrule
\end{tabular}
\caption{\textbf{Performance comparison of NeuroRisk against FauTE and the Gurobi optimum on \textsc{B4} and \textsc{IBM}.}}
\label{tab:faute_throughput}
\end{table}

\subsection{Runtime Analysis}
\label{subsec:eval_runtime}

We evaluate the scalability of NeuroRisk against solver-based methods along two critical dimensions: the number of failure scenarios ($N$) and the topology size.

\begin{figure}[t]
    \centering
    \includegraphics[width=0.85\columnwidth]{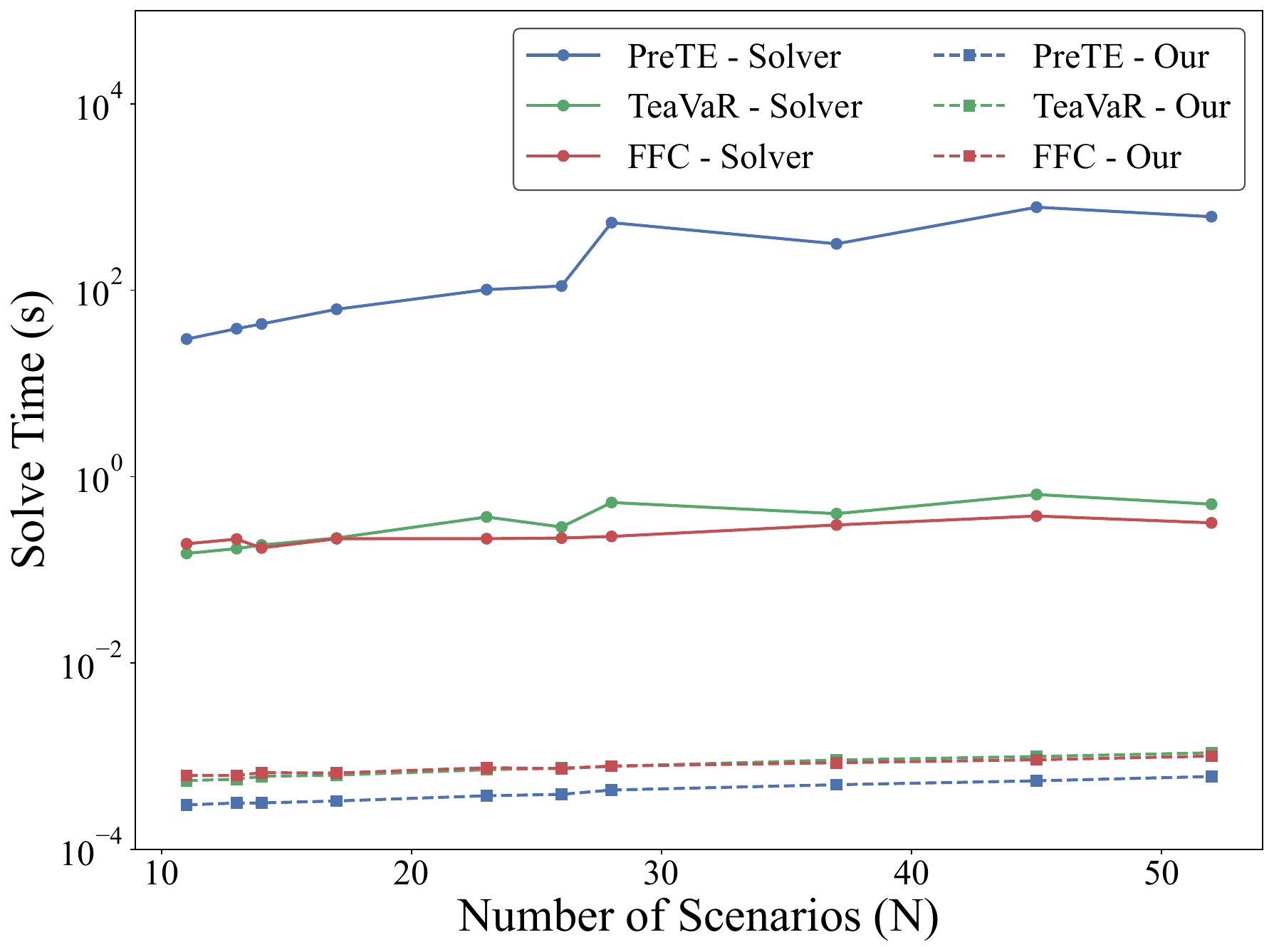}
    \caption{\textbf{Solve time vs.\ number of scenarios ($N$) on \textsc{IBM}.} 
    Solver-based methods (dashed lines) degrade significantly as $N$ grows, with PreTE spiking to $10^3$\,s due to decomposition overhead. In contrast, NeuroRisk (solid lines) maintains a flat, millisecond-level inference time.}
    \label{fig:eval_runtime}
\end{figure}

\noindent\textbf{Scaling with Scenarios ($N$).}
Figure~\ref{fig:eval_runtime} illustrates the impact of scenario count $N{=}|\mathcal{Q}|$ on solve time using the \textsc{IBM} topology.
Solver-based methods scale roughly linearly or super-linearly with $N$, because each additional scenario introduces a replicated block of constraints (and auxiliary variables) to the underlying LP formulation.
Notably, PreTE exhibits non-smooth runtime ``jumps'' (rising to $10^2$--$10^3$\,s), suggesting that the solver frequently enters harder combinatorial regimes (e.g., heavy branching or decomposition steps) once the formulation crosses certain complexity thresholds.
In contrast, NeuroRisk maintains a \textbf{flat scaling profile}, effectively decoupling inference latency from $N$ by parallelizing scenario processing within the neural architecture.

\noindent\textbf{Scaling with Topology.}
Importantly, the computational burden is not governed by $N$ alone, but by the \textbf{scenario-expanded problem size}, which scales with the product of $N$ and the routing decision space (number of tunnels/edges).
As shown in Table~\ref{tab:topology}, the number of tunnels grows rapidly from \textsc{IBM} to \textsc{GEANT} and \textsc{GERMANY50}.
Consequently, on these larger topologies, each additional scenario replicates constraints over thousands of tunnels, quickly pushing the solver beyond practical memory and iteration limits.
This explains the sharp transition observed in Table~\ref{tab:inference_latency}: while the solver handles \textsc{IBM} (${\sim}200$\,s), it consistently times out ($>$1\,h) on \textsc{GEANT} and \textsc{GERMANY50} due to this multiplicative explosion of the state space.
\begin{table}[b]
    \centering
    \small
    \setlength{\tabcolsep}{4pt}
    \begin{tabular}{@{}lrrr@{}}
    \toprule
    Topology & Solver (PreTE) & \multicolumn{2}{c}{NeuroRisk (Ours)} \\
    \cmidrule(l){3-4} 
     & (Time / Status) & \textbf{GPU} & \textbf{CPU (1-thread)} \\
    \midrule
    \textsc{B4} & 53.0\,s & 7\,ms & 151\,ms \\
    \textsc{IBM} & 196.4\,s & 11\,ms & 660\,ms \\
    \textsc{GEANT} & \textit{Timeout (>1h)} & 18\,ms & 7.0\,s \\
    \textsc{GERMANY50} & \textit{Timeout (>1h)} & 68\,ms & 177\,s \\
    \bottomrule
    \end{tabular}
    \caption{\textbf{Inference latency comparison.} For large topologies where the solver times out due to scenario-expanded constraint explosion, NeuroRisk remains feasible even on a single CPU thread. (NeuroRisk includes model construction time; $K{=}7$ unroll iterations).}
    \label{tab:inference_latency}
\end{table}

\noindent\textbf{Practical Deployment (Inference).}
In contrast to the solver's intractability on large graphs, NeuroRisk remains highly efficient for real-time deployment.
On a GPU, it solves the largest instances in under 70\,ms. 
Crucially, Table~\ref{tab:inference_latency} demonstrates that even on a \textbf{single CPU thread}, NeuroRisk produces feasible solutions for \textsc{GEANT} in just 7 seconds.
This confirms that our lightweight model is deployable on commodity control plane hardware without requiring specialized accelerators.

\noindent\textbf{Training Overhead.}
While NeuroRisk requires offline training, this is a manageable one-time cost.
Even for \textsc{GERMANY50}, training converges in under \textbf{4\,hours} on a single GPU (see Appendix~\ref{app:training_convergence}). 
Given that training occurs infrequently (e.g., only upon major topology changes), this offline overhead is negligible compared to the continuous, millisecond-level efficiency gains during online operation.

\subsection{Generalization}
\label{subsec:eval_generalization}

\begin{figure*}[t]
\centering
\begin{minipage}{0.5\textwidth}
    \centering
    \includegraphics[width=\textwidth]{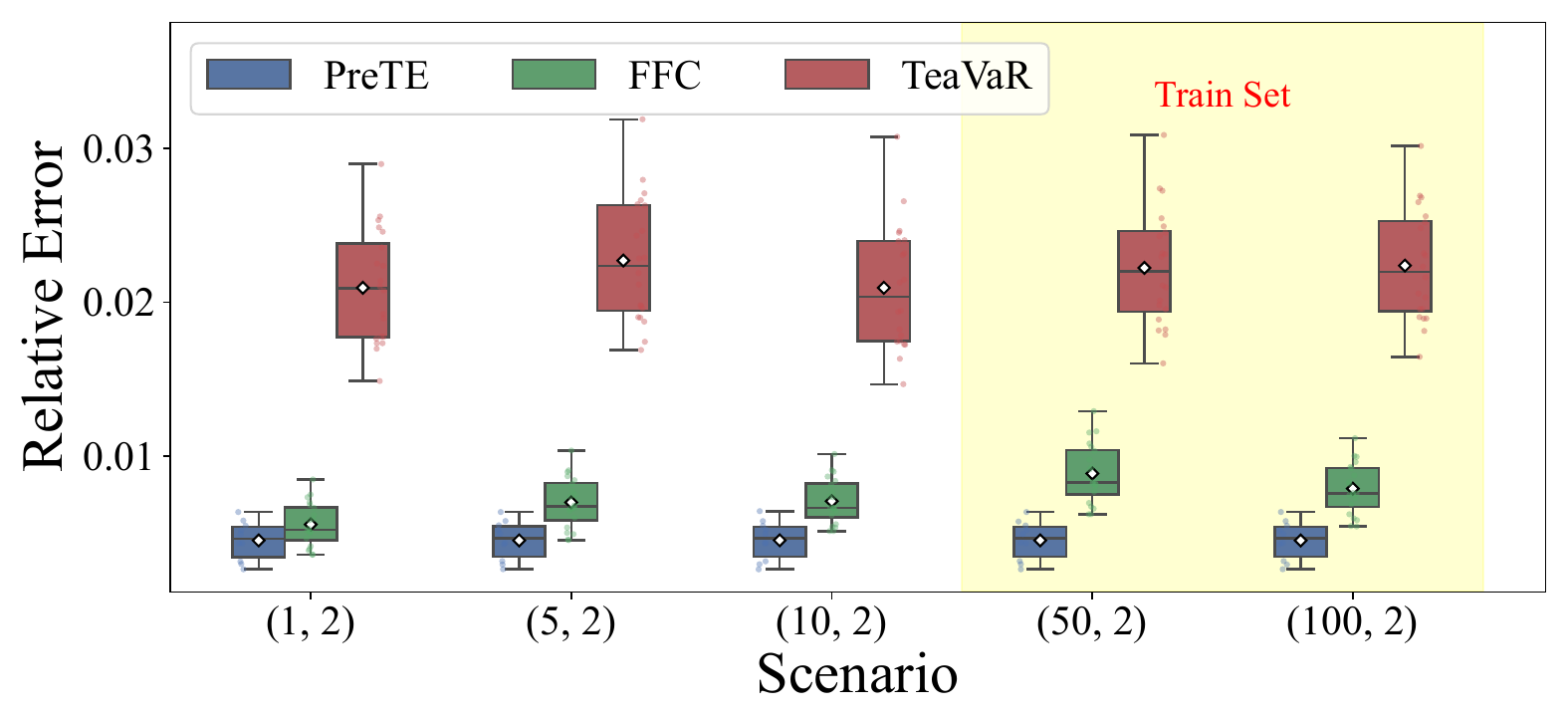}
    \vspace{-0.6em}
    {\small\textbf{(a)} Scenario distribution shift.}
\end{minipage}\hfill
\begin{minipage}{0.5\textwidth}
    \centering
    \includegraphics[width=\textwidth]{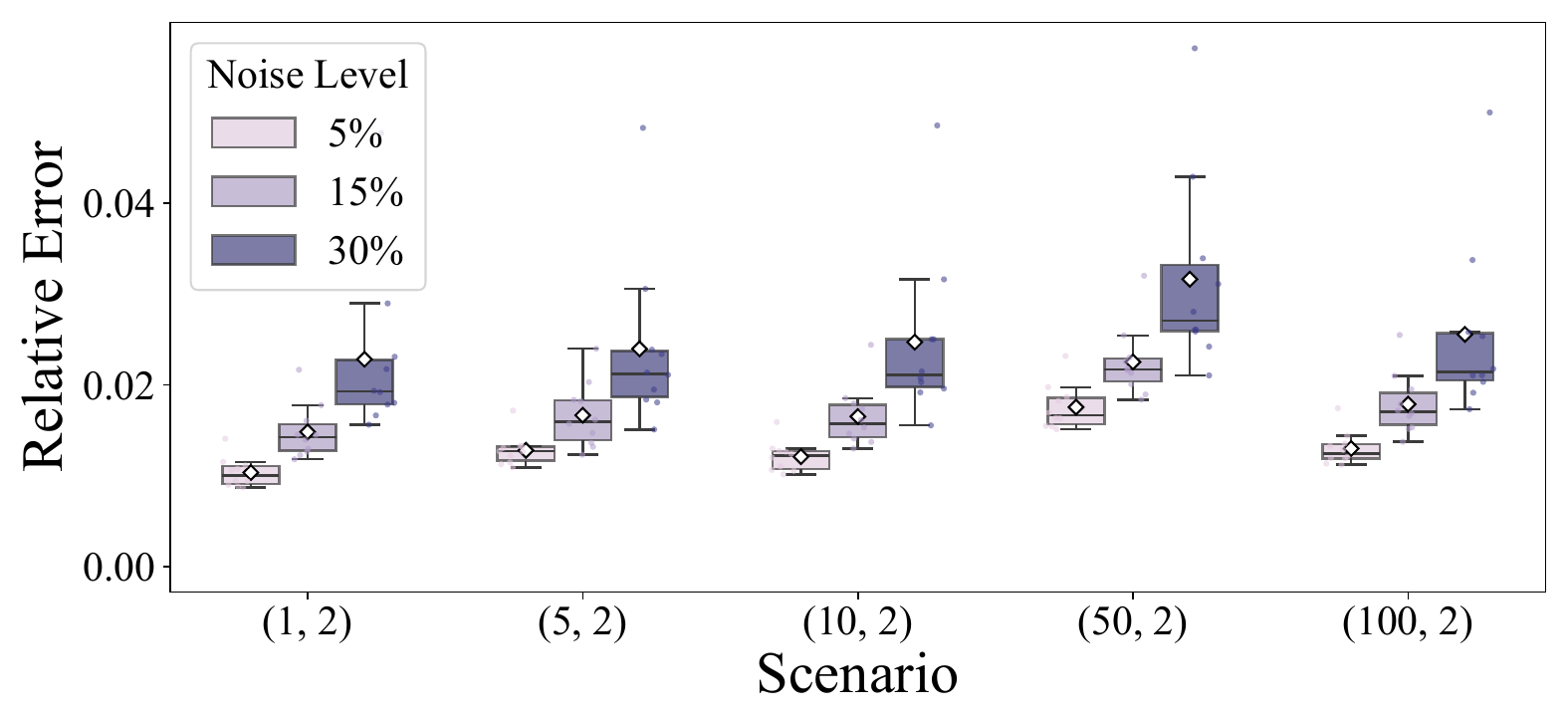}
    \vspace{-0.6em}
    {\small\textbf{(b)} Demand perturbation.}
\end{minipage}

\vspace{0.6em}

\begin{minipage}{0.5\textwidth}
    \centering
    \includegraphics[width=\textwidth]{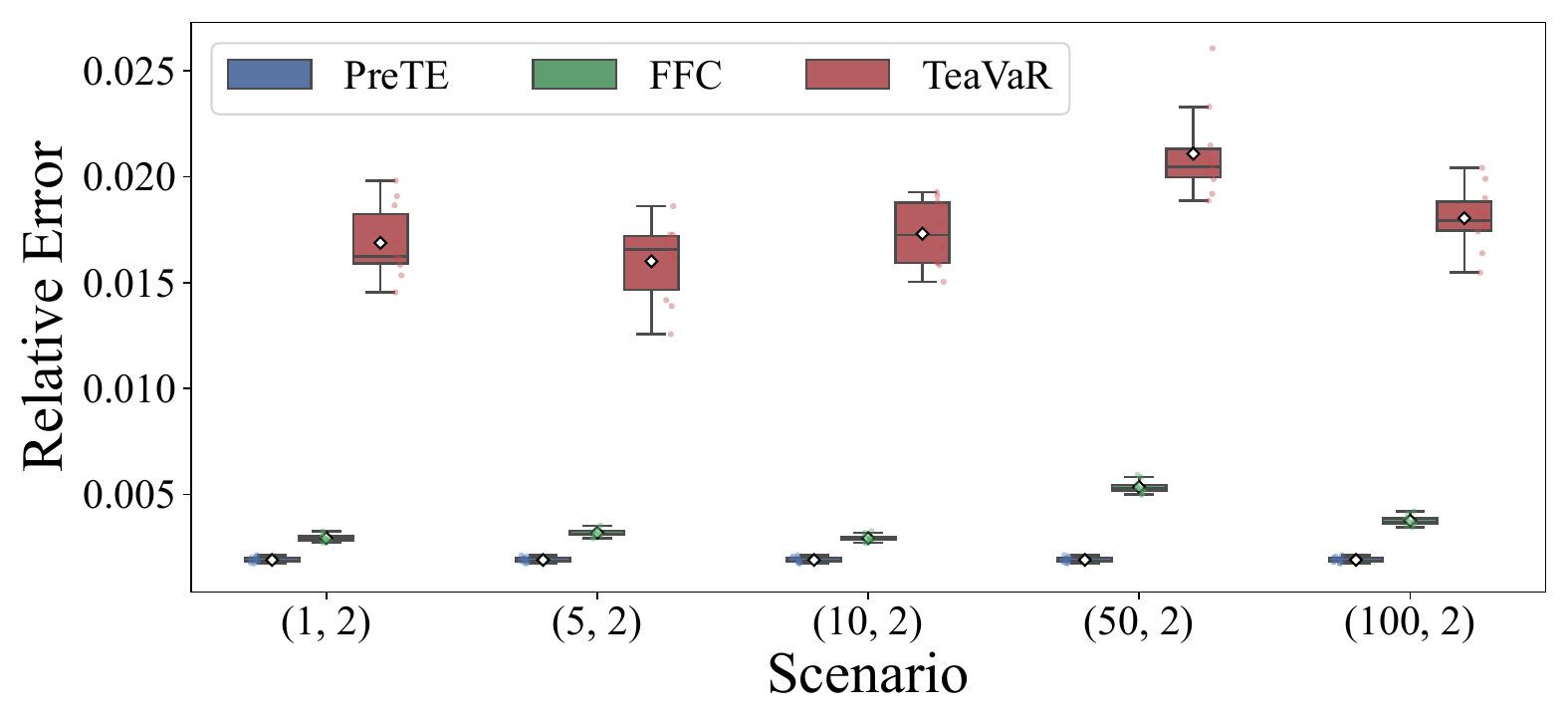}
    \vspace{-0.6em}
    {\small\textbf{(c)} Tunnel-set shift.}
\end{minipage}\hfill
\begin{minipage}{0.50\textwidth}
    \centering
    \includegraphics[width=\textwidth]{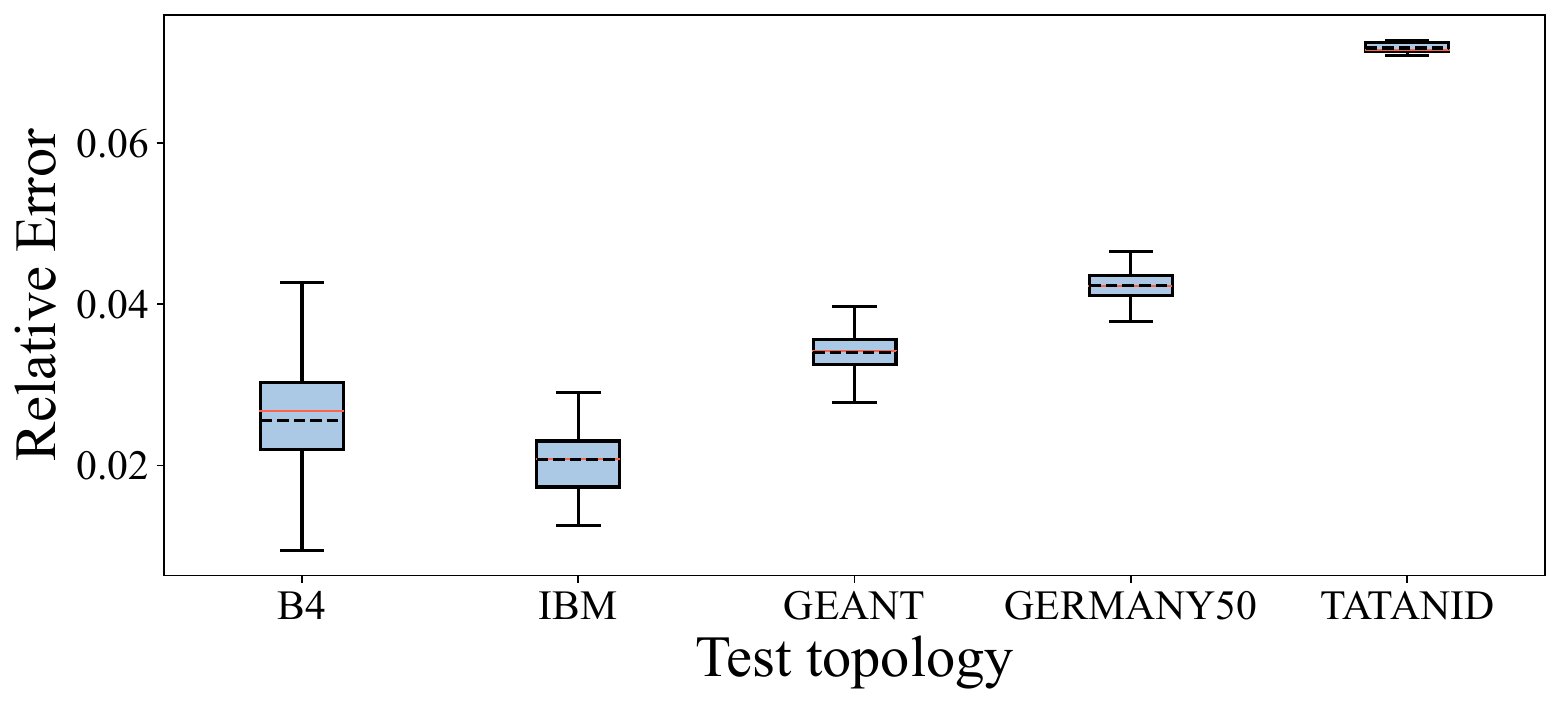}
    \vspace{-0.6em}
    {\small\textbf{(d)} Cross-topology (TeaVaR).}
\end{minipage}

\caption{\textbf{Generalization summary.} NeuroRisk generalizes across multiple distribution shifts without retraining: (a) unseen scenario configurations $(c,s)$ on \textsc{IBM}, (b) demand-proportional perturbations on \textsc{B4} (TeaVaR), (c) changes to candidate tunnel sets on \textsc{B4}, and (d) zero-shot transfer to unseen topologies for the TeaVaR objective. For readability, subfigures (a)--(d) show cropped views; Appendix~\ref{app:full_generalization_experiment} provides the full, uncropped plots and additional variants across objectives and topologies.}
\label{fig:generalization_summary}
\end{figure*}

\noindent\textbf{Generalization across scenario distributions.}
\label{subsec:eval_scenario_generalization}
We first evaluate scenario-distribution generalization by testing on unseen scenario configurations $(c,s)$ (different cutoff thresholds and Weibull scales) while keeping model weights fixed.
Figure~\ref{fig:generalization_summary}(a) shows a representative example on \textsc{IBM}.
Across objectives, NeuroRisk generalizes well to unseen test configurations (e.g., $(1, 2)$ and $(5, 2)$), indicating robustness to changes in scenario pruning and failure severity.

\noindent\textbf{Robustness to demand perturbations.}
\label{subsec:eval_noise}
To test robustness to traffic dynamics, we perturb the input demand by injecting zero-mean Gaussian noise whose standard deviation is proportional to the original demand and then measure the resulting change in relative error.
Importantly, the training set and model weights remain fixed: we do not retrain or fine-tune on perturbed data. The perturbation is applied only to the test instances at inference time, evaluating out-of-distribution generalization under demand shifts.
Concretely, for each demand $D_{f}$, we sample a noise term $\eta_{f} \sim \mathcal{N}\left(0,\left(\sigma D_{f}\right)^{2}\right)$ and form the perturbed demand $D_{f}^{\prime}=\max \left\{0, D_{f}+\eta_{f}\right\}$, where the truncation enforces the physical constraint that demands are nonnegative. We evaluate three noise levels $\sigma \in\{5 \%, 15 \%, 30 \%\}$, corresponding to standard deviations of $0.05 D_{f}, 0.15 D_{f}$, and $0.30 D_{f}$, respectively.

Figure~\ref{fig:generalization_summary}(b) shows the effect of demand perturbations on \textsc{B4} with the TeaVaR objective. Even under 30\% demand noise, NeuroRisk maintains relative errors below 5\% in most configurations.
The degradation is graceful: the relative error increases approximately linearly with the noise level, indicating stable gradient propagation through the Gated Reservation mechanism.
This robustness is crucial for practical deployment, where runtime traffic matrices inevitably deviate from the historical distributions used during training.

\noindent\textbf{Generalization across tunnel configurations.}
\label{subsec:eval_ksp}
NeuroRisk is trained with KSP candidate sets of size $K_{sp}{=}3$ per source-destination pair.
To assess robustness to changes in the action space, we perform a \emph{cross-$K_{sp}$} evaluation: we keep the trained model fixed and test on instances whose candidate tunnel sets are generated with $K_{sp}{=}2$.  
Figure~\ref{fig:generalization_summary}(c) shows that NeuroRisk remains accurate under this tunnel set shift. This robustness follows from the edge-centric reservation parameterization: since each edge's Softmax operates over whichever tunnels traverse it, adding or removing tunnels simply changes the Softmax dimension without requiring architectural modifications.
This flexibility is valuable for operators who may dynamically adjust routing options according to conditions.

\noindent\textbf{Cross-topology generalization.}
\label{subsec:eval_topology_generalization}
We next evaluate whether a \emph{single} TeaVaR model can generalize across \emph{unseen} network topologies.
We train one model on a mixture of \textsc{IBM} and \textsc{GEANT} instances (with unified padding and masking to support multi-topology batching), select the checkpoint using validation instances from the same two training topologies, and then test \emph{zero-shot} on held-out instances from \textsc{IBM}/\textsc{GEANT} as well as unseen \textsc{B4}/\textsc{GERMANY50}/\textsc{TATANID}. Throughout, we treat \textsc{B4}, \textsc{GERMANY50}, and \textsc{TATANID} as \emph{pure} target domains: we perform no retraining or fine-tuning on these topologies.
This isolates cross-topology transfer and avoids per-topology training overhead, which can be expensive on large graphs (e.g., \textsc{TATANID}).
Figure~\ref{fig:generalization_summary}(d) summarizes the relative error compared to the optimum in each test topology.
The gap in unseen \textsc{B4} is minimal, and even on the much larger \textsc{TATANID}, relative error remains low ($<8\%$), confirming robust transfer across diverse scales.

\noindent\textbf{Generalization across iteration counts.}
\label{subsec:eval_iter_generalization}
Because shared weights are reused at every unroll step, a model trained at $K{=}7$ can be deployed at any $K_1 \le K$ without retraining, allowing operators to trade solution quality for latency at runtime.
On \textsc{B4} (TeaVaR), relative error degrades gracefully from $1.35\%$ ($K_1{=}7$) to $7.72\%$ ($K_1{=}1$), with latency scaling linearly in $K_1$ (Appendix~\ref{app:iter_sensitivity}).

\subsection{Ablation and Module Analysis}
\label{subsec:ablation}

\noindent\textbf{Feasibility enforcement comparison.} We conduct an ablation study to evaluate the effectiveness of our Gated Reservation (GR) mechanism compared to other feasibility enforcement methods.
Specifically, we compare:
(1)  {Global Scaling (GS)}: Post-hoc scaling by a network-wide factor $\gamma_{\max}$ (\S\ref{sec:challenges}).
(2)  {Local Scaling (LS)}: Post-hoc scaling by per-tunnel factors $\gamma_t$ (\S\ref{subsec:gated_reservation}).
(3)  {Bandwidth Reservation (BR)}: Per-edge Softmax without tunnel-level gating (Eq.~\eqref{eq:naive_softmax}).
(4)  {Gated Reservation (GR)}: Our full mechanism with tunnel-level gating (Eq.~\eqref{eq:gated_softmax}).
\begin{figure}[thb]
\centering
\includegraphics[width=\columnwidth]{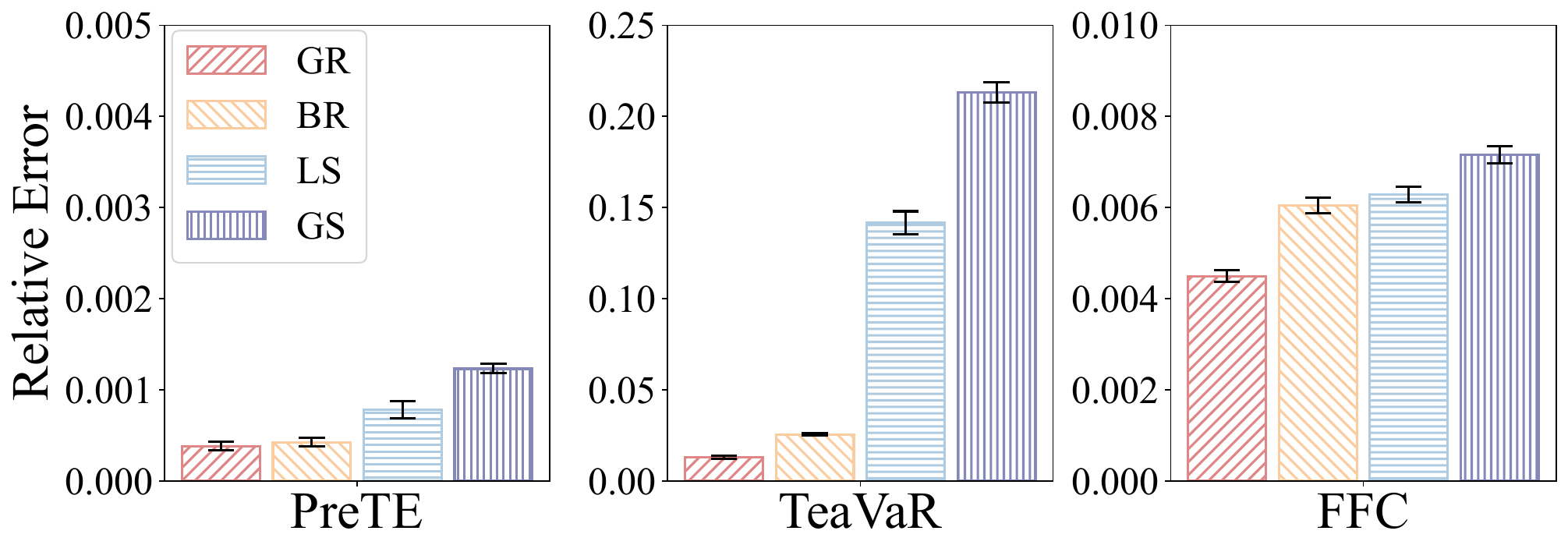}
\caption{\textbf{Ablation study on \textsc{B4}.} Comparison of feasibility enforcement methods (GS, LS, BR, GR) across different objectives. GR consistently achieves the lowest relative error by combining construction-based feasibility with tunnel-level coordination.}
\label{fig:ablation}
\end{figure}
Figure~\ref{fig:ablation} presents the results on \textsc{B4}.
GS and LS exhibit significantly higher relative errors due to the optimization fragility and projection discontinuities discussed in \S\ref{sec:challenges}.
BR (Eq.~\eqref{eq:naive_softmax}) eliminates scaling pitfalls but suffers from the Short-Board Trap (\S\ref{subsec:module1}), as it lacks a mechanism for tunnel-level coordination.
The GR mechanism (Eq.~\eqref{eq:gated_softmax}) achieves near-optimal performance by restoring tunnel-level coordination within a smooth, feasibility-by-construction framework.
The results confirm that both construction-based feasibility and tunnel-level gating are essential.

\noindent\textbf{Route diversity analysis.}
As hypothesized in \S\ref{subsec:gated_reservation}, the LS approach exhibits a systemic bias toward short tunnels. We validate this by measuring the \emph{direct ratio}---the fraction of carried traffic allocated to 1-hop tunnels---on \textsc{B4}. As shown in Table~\ref{tab:teavar_route_structure} (Appendix~\ref{app:ablation_diagnostics}), LS yields a direct ratio of 0.989, compared to 0.883 for GR, confirming that LS collapses route diversity toward direct tunnels.

\noindent\textbf{Nominal (no-failure) comparison.}
To demonstrate that GR's benefits extend beyond failure scenarios, we evaluate throughput under the nominal (failure-free) setting. The detailed results, summarized in Table~\ref{tab:maxflow_nominal} (Appendix~\ref{app:ablation_diagnostics}), show that GR achieves 97\% of optimal throughput on \textsc{GEANT} and 96\% on \textsc{B4}, whereas the widely used GS method yields only 13\% on \textsc{GEANT} and 17\% on \textsc{B4}.
These results underscore that GR's feasibility-by-construction design is effective for general TE, not only for risk-aware objectives.

\section{Discussion}
\label{sec:discussion}

\noindent\textbf{Deployment flexibility.}
As shown in \S\ref{subsec:eval_iter_generalization}, operators can reduce the number of unroll iterations at deployment time to trade solution quality for latency without retraining. Only training requires a GPU; inference runs efficiently on CPU (\S\ref{subsec:eval_runtime}), enabling deployment on SDN controllers or edge servers without dedicated accelerators. For unseen topologies, an existing model can serve as a warm-start (\S\ref{subsec:eval_topology_generalization}); once a topology-specific model is trained, it can be swapped in.

\noindent\textbf{Scalability to large scenario sets.}
The \emph{Sort-and-Select} formulation decouples scenarios: loss and gradient computation proceed independently per scenario, then aggregate via weighted summation (\S\ref{subsec:perception}). During training, we leverage gradient checkpointing to avoid storing intermediate computation graphs across scenarios; this makes memory consumption independent of $|\mathcal{Q}|$, so increasing the scenario count only affects training time. When exhaustive enumeration is infeasible (e.g., $\hat{k}$-failure models with $O(\binom{|E|}{\hat{k}})$ scenarios), practitioners can prune to high-probability or high-impact scenarios without altering the algorithm.

\section{Related Work}
\label{sec:related}

\noindent\textbf{TE Optimization and Demand Robustness.}
Centralized TE optimizes utilization via global programs (e.g., SWAN~\cite{swan} and B4~\cite{b4_after}), while scalability motivates decomposition methods such as NCFlow~\cite{ncflow} and SSDO~\cite{ssdo}.
Recent learning-based systems further accelerate TE optimization and improve generalization (e.g., Teal~\cite{xu_teal_2023}, HARP~\cite{harp}, RedTE~\cite{redte}).
Robustness to demand uncertainty has been studied via worst-case formulations (e.g., Oblivious Routing~\cite{Oblivious_Routing} and COPE~\cite{cope}) and semi-oblivious routing (e.g., SMORE/Semi-Oblivious~\cite{Semi-Oblivious}); complementary approaches incorporate prediction errors or burstiness more explicitly (e.g., Figret~\cite{figret}).

\noindent\textbf{Failure Robustness and Risk-Aware TE.}
Proactive methods plan across failure scenarios to enforce availability or tail-risk objectives (e.g., FFC~\cite{ffc}, PCF~\cite{pcf}, TeaVaR~\cite{teavar}), while complementary systems emphasize fast reaction and cross-layer restoration (e.g., Flexile~\cite{flexile}, Arrow~\cite{ARROW}, PreTE~\cite{prete}).
FauTE~\cite{faute} applies deep learning with a probability-weighted penalty; in contrast, NeuroRisk enforces feasibility by construction and optimizes scenario-aware risk objectives, avoiding scaling/projection pitfalls. Moreover, NeuroRisk naturally extends to demand robustness: by treating distinct traffic matrices as scenarios in $\mathcal{Q}$, the same architecture can optimize over demand uncertainty without modification.

\section{Conclusion}
Risk-aware traffic engineering must jointly handle expressive objectives beyond MLU, strict capacity feasibility, and a probabilistic, variable-sized failure scenario set as a runtime input. We propose NeuroRisk, a physics-informed neural optimizer that achieves feasibility by construction via Gated Reservation (edge-local reservations with tunnel-level coordination) and supports general \emph{Sort-and-Select} risk objectives using a physics-aware update network that is invariant to scenario ordering. Across WAN topologies and risk formulations, NeuroRisk matches solver solution quality with orders-of-magnitude speedups, and it generalizes to demand shifts and candidate tunnel changes without retraining; future work will extend evaluation to additional topologies and traffic sources and further integrate online forecast updates. Our implementation will be made available on GitHub to support further research in AI-driven network optimization. \textbf{This work does not raise any ethical concerns.} 

\newpage
\bibliographystyle{ACM-Reference-Format}
\bibliography{reference}

\newpage
\appendix

\section*{Appendix}
\section{Full Notation}
\label{app:notation}

To facilitate a deeper understanding of the NeuroRisk framework, we provide a comprehensive summary of the notation used throughout this paper in Table~\ref{tab:notation_full}. Our notation is organized into four logical groups:
(i) \textbf{Network Topology and Traffic:} Variables such as $\mathcal{G}$, $C_e$, and $D_f$ describe the underlying physical graph and the input traffic demands.
(ii) \textbf{Routing and Allocation Decisions:} $x_{f,t}$ and $y_{t,e}$ represent the target traffic fractions and link-level reservation ratios, respectively, which are the primary decision variables optimized by our framework.
(iii) \textbf{Failure Scenarios and Risk Metrics:} To characterize network reliability, we use $q$ and $p_q$ to denote probabilistic failure scenarios, and $\ell_{f,q}$ to capture the resulting loss ratio. The variables $v_{f,r}$ and $\pi_{f,r}$ are specifically defined for the ranking-based risk assessment (e.g., CVaR computation).
(iv) \textbf{Neural Optimization Components:} $\Phi_{\text{phy}}$ and $\Psi_{\theta}$ denote the core functional blocks of our unrolled optimization architecture, responsible for physics-informed feature extraction and iterative state updates.

\begin{table}[htbp]
    \centering
    \small
    \begin{tabular}{@{}lp{0.72\columnwidth}@{}}
        \toprule
        Symbol & Meaning \\
        \midrule
        $\mathcal{G}(\mathcal{V}, \mathcal{E})$ & Network graph with nodes $\mathcal{V}$ and links $\mathcal{E}$. \\
        $f \in \mathcal{F}$ & Traffic demand (flow). \\
        $\mathcal{T}_f$ & Candidate tunnels for flow $f$. \\
        $\mathcal{T}_e$ & The set of tunnels traversing edge $e$. \\
        $q \in \mathcal{Q}$, $p_q$ & Failure scenario and its probability. \\
        $C_e$ & Capacity of link $e$. \\
        $D_f$, $\mathbf{D}$ & Demand of flow $f$ and the demand vector. \\
        $x_{f,t}$, $\mathbf{x}$ & Target fraction of $D_f$ to be routed on tunnel $t$ (allocation decision). \\
        $\alpha_{t,q}$ & Indicator that tunnel $t$ survives in scenario $q$. (1 if tunnel $t$ survives in $q$, 0 otherwise) \\
        $\ell_{f,q}$ & Realized loss ratio of flow $f$ under scenario $q$. \\
        $v_{f,r}$ & $r$-th largest loss for flow $f$ after sorting scenarios. \\
        $\pi_{f,r}$ & Probability mass of the scenario ranked $r$ for flow $f$. \\
        $\Gamma_{f,r}$ & Cumulative probability up to rank $r$: $\sum_{j=1}^r \pi_{f,j}$. \\
        $I_{f,r}$ & Binary selection mask encoding the risk preference. \\
        $m_q$ & Scenario-level selection mask (e.g., $1$ for tail scenarios in CVaR, $0$ otherwise). \\
        $\rho_q$ & Scenario risk weight: $\rho_q = p_q \cdot m_q$. \\
        $\delta_{f,q,r}$ & Indicator that scenario $q$ is ranked $r$ for flow $f$. \\
        $z_{t,e}$, $y_{t,e}$ & Latent contention logit and resulting reservation ratio on link $e$ for tunnel $t$. \\
        $w_t$ &  Tunnel-wide priority score shared across all edges of $t$.\\
        $\Phi_{\text{phy}}$, $\Psi_{\theta}$ & Physics feature extractor and learned update network. \\
        \bottomrule
    \end{tabular}
        \caption{Notation used throughout the paper.}
        \label{tab:notation_full}

\end{table}

\section{Hyperparameter Settings}
\label{app:hyperparams}

Table~\ref{tab:hyperparams} summarizes the representative hyperparameter configuration for our model. These settings were determined via a systematic \textbf{grid search} covering learning rate $\eta \in \{10^{-4}, 5\times10^{-4}, 10^{-3}\}$, hidden layer dimensions $\in \{32, 64, 128\}$, and the number of unrolled iterations $K$. The selected $K=7$ strikes an optimal balance between the quality of the optimization solution and inference latency.

The update network $\Psi_{\theta}$ is implemented as a 2-layer MLP to ensure minimal per-iteration computational overhead. We employ the Adam optimizer with an \textbf{early stopping} patience of 10 epochs, which provides sufficient training stability across various network scales. 

To manage the memory footprint during the processing of a large number of failure scenarios $|\mathcal{Q}|$, our framework includes an optional \textbf{scenario chunking} mechanism. This feature is disabled by default to maximize throughput on standard benchmarks. For large-scale topologies, we enable chunking with a size ranging from \textbf{1 to 20}, depending on the available GPU VRAM and the specific complexity of the topology. This mechanism ensures that the training and inference remain feasible on a single commercial GPU without altering the gradient integrity.

\begin{table}[htbp]
    \centering
    \small
    \begin{tabular}{@{}ll@{}}
        \toprule
        Hyperparameter & Value \\
        \midrule
        Architecture & Unrolled MLP-based Optimization \\
        Update MLP & 2-layer MLP (ReLU) \\
        Hidden dimension & 64 \\
        Input dimension & 8 \\
        Unroll iterations $K$ & 7 \\
        Training epochs & 30 \\
        Early stopping patience & 10 \\
        Batch size & 16 \\
        Learning rate & $10^{-3}$ \\
        Scenario chunking & Disabled by default (1--20 for large) \\
        \bottomrule
    \end{tabular}
        \caption{Key hyperparameters.}
        \label{tab:hyperparams}
\end{table}

\section{Derivation of Physics Engine Features}
\label{app:physics_features}

This appendix derives why the eight features in Table~\ref{tab:features} suffice for the update network $\Psi_\theta$ to produce effective increments $(\Delta\mathbf{Z},\Delta\mathbf{W})$. The argument follows the chain rule for $\partial \mathcal{J}_q / \partial z_{t,e}$ and $\partial \mathcal{J}_q / \partial w_t$ (Eq.~\eqref{eq:micro_gradient} in \S\ref{subsec:perception}).

\noindent\textbf{Risk term $\partial \mathcal{J}_q / \partial x_{f,t}$.}
The scenario objective $\mathcal{J}_q$ is a function of per-flow losses $\ell_{f,q}$. For flow $f$ and tunnel $t \in \mathcal{T}_f$, $\ell_{f,q} = \max\bigl(0, 1 - \sum_{t'} x_{f,t'} \alpha_{t',q}\bigr)$, so $\partial \mathcal{J}_q / \partial x_{f,t}$ depends on: (i) whether tunnel $t$ survives in $q$, i.e.\ $\alpha_{t,q}$; (ii) the flow's loss under $q$, i.e.\ $\ell_{f,q}$; and (iii) the scenario-level loss $\ell_q$ (e.g., for CVaR-style masks). Hence the \textbf{Risk} features: $\alpha_{t,q}$, $\ell_{f,q}$, $\ell_q$.

\noindent\textbf{Physics term $\partial x_{f,t} / \partial y_{t,e}$.}
We have $x_{f,t} = b_t / D_{f(t)}$ with $b_t = \min_{e' \in t} C_{e'} y_{t,e'}$. So $x_{f,t}$ is determined by the bottleneck edge along tunnel $t$. The derivative $\partial x_{f,t} / \partial y_{t,e}$ is nonzero only when edge $e$ is the bottleneck for $t$ (i.e., $C_e y_{t,e} = b_t$). It depends on: (i) the current satisfaction $x_t = b_t / D_{f(t)}$; (ii) the \emph{bottleneck margin} $m_{t,e} = C_e y_{t,e} - b_t$ (zero at the bottleneck, positive elsewhere); (iii) capacity $C_e$ and demand $D_{f(t)}$ (scale the derivative). Hence the \textbf{Physics} features: $x_t$, $m_{t,e}$, $C_e$, $D_{f(t)}$.

\noindent\textbf{Saturation term $\partial y_{t,e} / \partial z_{t,e}$.}
Under the gated Softmax $y_{t,e} = \exp(z_{t,e}+w_t) / \sum_{k \in \mathcal{T}_e} \exp(z_{k,e}+w_k)$, we have $\partial y_{t,e} / \partial z_{t,e} = y_{t,e}(1 - y_{t,e})$. So the sensitivity of the reservation to the logit is determined by the current $y_{t,e}$. Hence the \textbf{Saturation} feature: $y_{t,e}$.

\noindent\textbf{Summary.}
The chain rule thus identifies exactly the eight quantities in Table~\ref{tab:features}: three Risk, four Physics, and one Saturation. Feeding these as $\mathbf{s}_{t,e,q}$ allows $\Psi_\theta$ to approximate the effect of an update on the risk objective without computing gradients explicitly.

\section{Scenario Set Statistics}
\label{app:scenario_stats}

\begin{table}[thb]
\centering
\small
\setlength{\tabcolsep}{4.0pt}
\renewcommand{\arraystretch}{1.15}
\begin{tabular}{@{}llrrrrr@{}}
\toprule
Topology & $s$ & $c{=}1$ & $c{=}5$ & $c{=}10$ & $c{=}50$ & $c{=}100$ \\
\midrule
\multirow{2}{*}{\textsc{B4}}
  & 2 & 33  & 22  & 19  & 10  & 10  \\
  & 4 & 34  & 34  & 22  & 16  & 11  \\
\midrule
\multirow{2}{*}{\textsc{GEANT}}
  & 2 & 80  & 40  & 33  & 23  & 18  \\
  & 4 & 239 & 108 & 46  & 27  & 23  \\
\midrule
\multirow{2}{*}{\textsc{GERMANY50}}
  & 2 & 242 & 107 & 76  & 58  & 39  \\
  & 4 & 671 & 294 & 134 & 68  & 53  \\
\midrule
\multirow{2}{*}{\textsc{IBM}}
  & 2 & 37  & 26  & 23  & 14  & 11  \\
  & 4 & 52  & 45  & 28  & 17  & 13  \\
\midrule
\multirow{2}{*}{\textsc{TATANID}}
  & 2 & 871 & 227 & 159 & 107 & 72  \\
  & 4 & 2787& 551 & 236 & 127 & 100 \\
\bottomrule
\end{tabular}
\caption{\textbf{Scenario counts after pruning (sorted by $c$).} Number of retained scenarios $|\mathcal{Q}|$ for each topology. }
\label{tab:scenario_counts}
\end{table}
Table~\ref{tab:scenario_counts} indicates that $|\mathcal{Q}|$ is driven by (i) topology scale, which increases the number of correlated multi-link failure combinations, and (ii) scenario pruning/severity $(c,s)$, which controls how much tail mass survives. Across all topologies, smaller $c$ (weaker cutoff) monotonically increases $|\mathcal{Q}|$ by admitting more low-probability combinations; larger $s$ (heavier tail) further increases $|\mathcal{Q}|$ by pushing more multi-link events above the cutoff. 
This scaling causally explains solver runtime jumps: scenario-enumerating baselines expand in constraints/subproblems with $|\mathcal{Q}|$, so larger graphs or heavier-tail $(c,s)$ push the solver into harder decomposition/branching regimes.

\section{Generalization Results on Additional Topologies and Objectives}
\label{app:full_generalization_experiment}

This appendix complements \S\ref{subsec:eval_generalization} by providing the full, uncropped versions of the plots that are summarized in Figure~\ref{fig:generalization_summary}. In the main paper, we crop subfigures to improve readability under tight space constraints and to keep a consistent visual scale across panels; here we report the complete figures and additional variants across objectives and topologies for completeness and reproducibility.
\begin{figure}[h]
\centering
\includegraphics[width=\columnwidth]{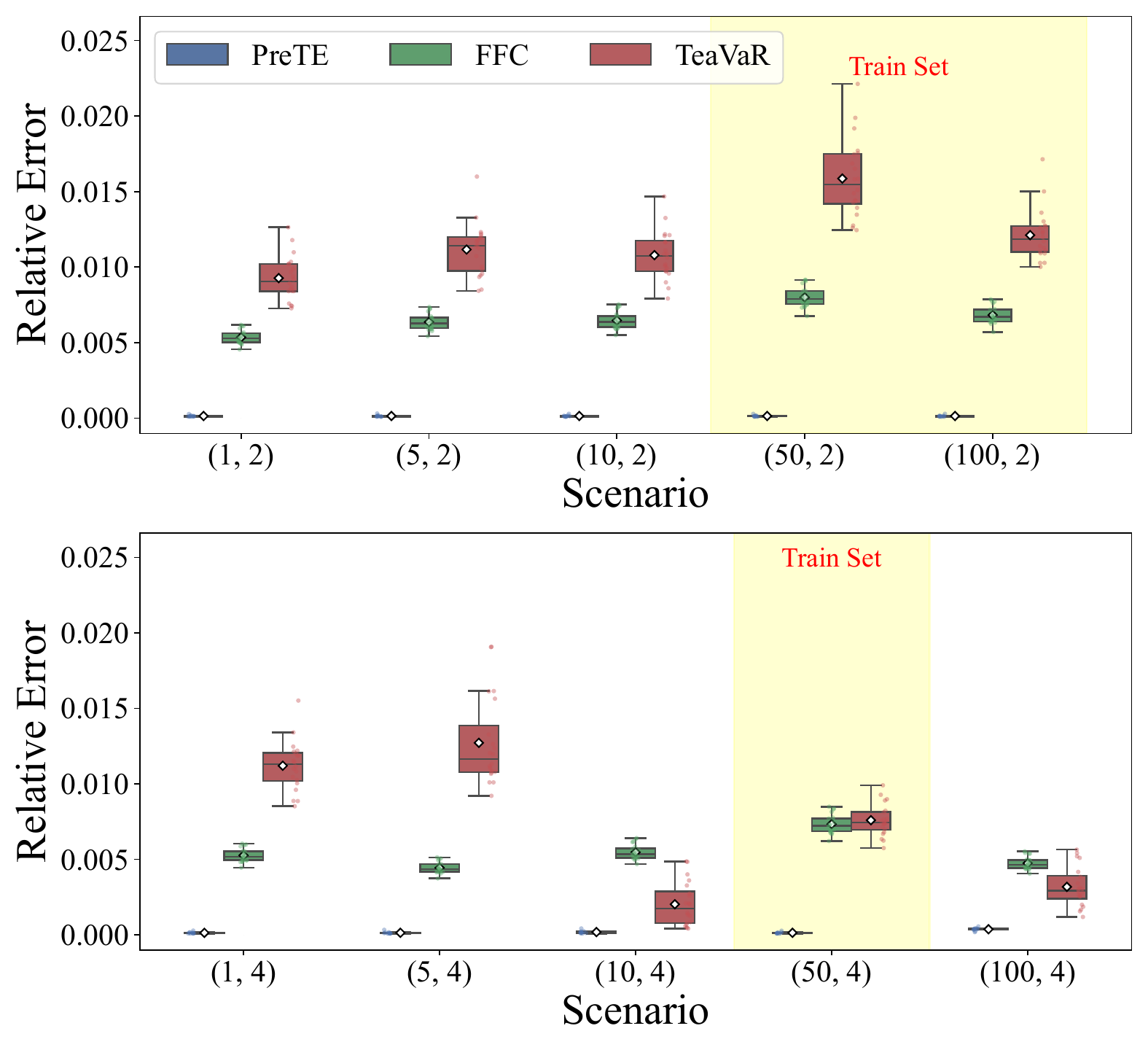}
\caption{\textbf{Scenario-distribution generalization on \textsc{B4}.} Relative error against solver baselines (FFC/PreTE/TeaVaR) across scenario configurations. Bars marked with yellow background indicate training configurations.}
\label{fig:app_b4_relerr}
\end{figure}
\begin{figure}[h]
\centering
\includegraphics[width=\columnwidth]{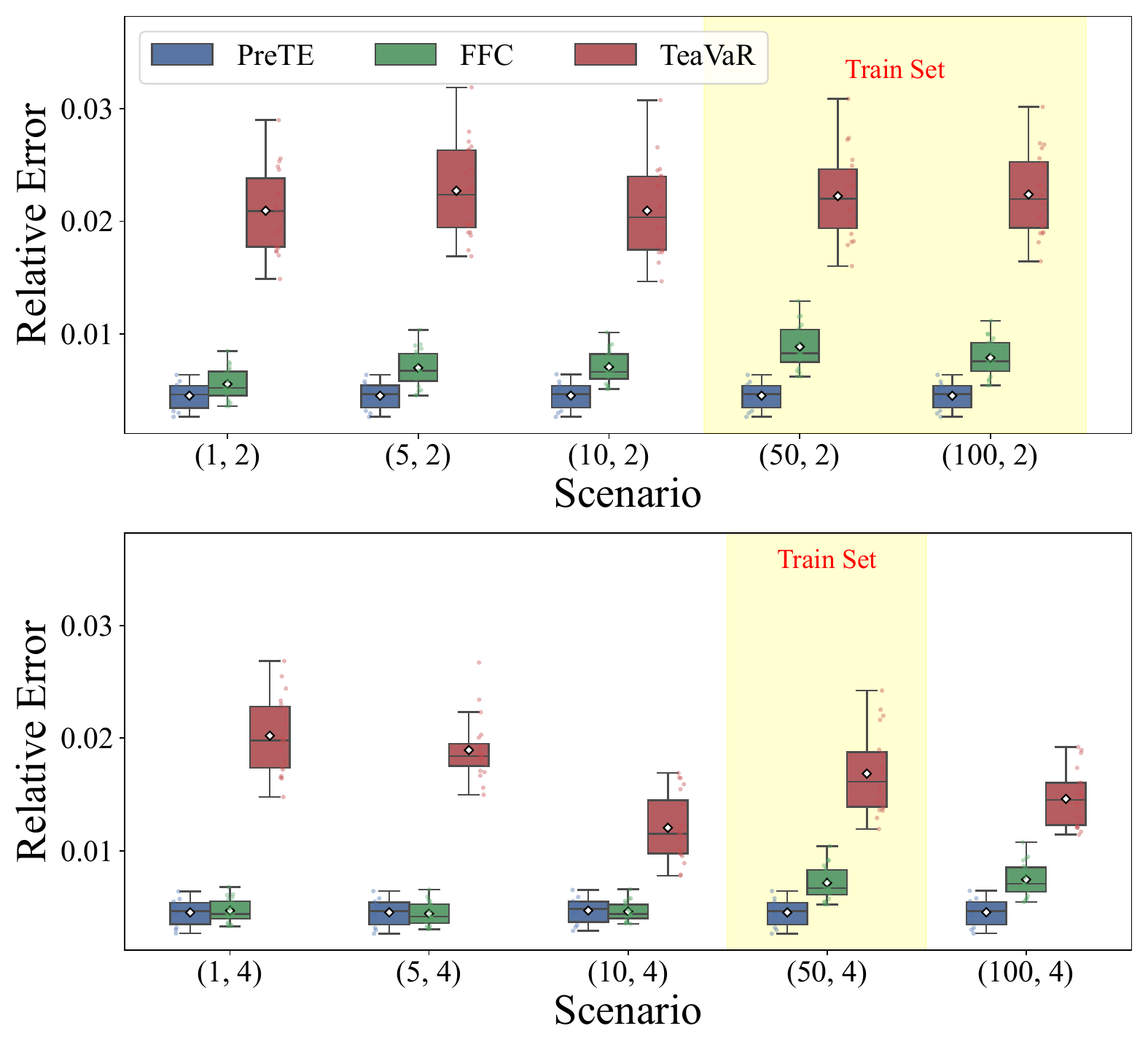}
\caption{\textbf{Scenario-distribution generalization on \textsc{IBM}.} Same as above.}
\label{fig:app_ibm_relerr}
\end{figure}
\begin{figure}[h]
\centering
\includegraphics[width=\columnwidth]{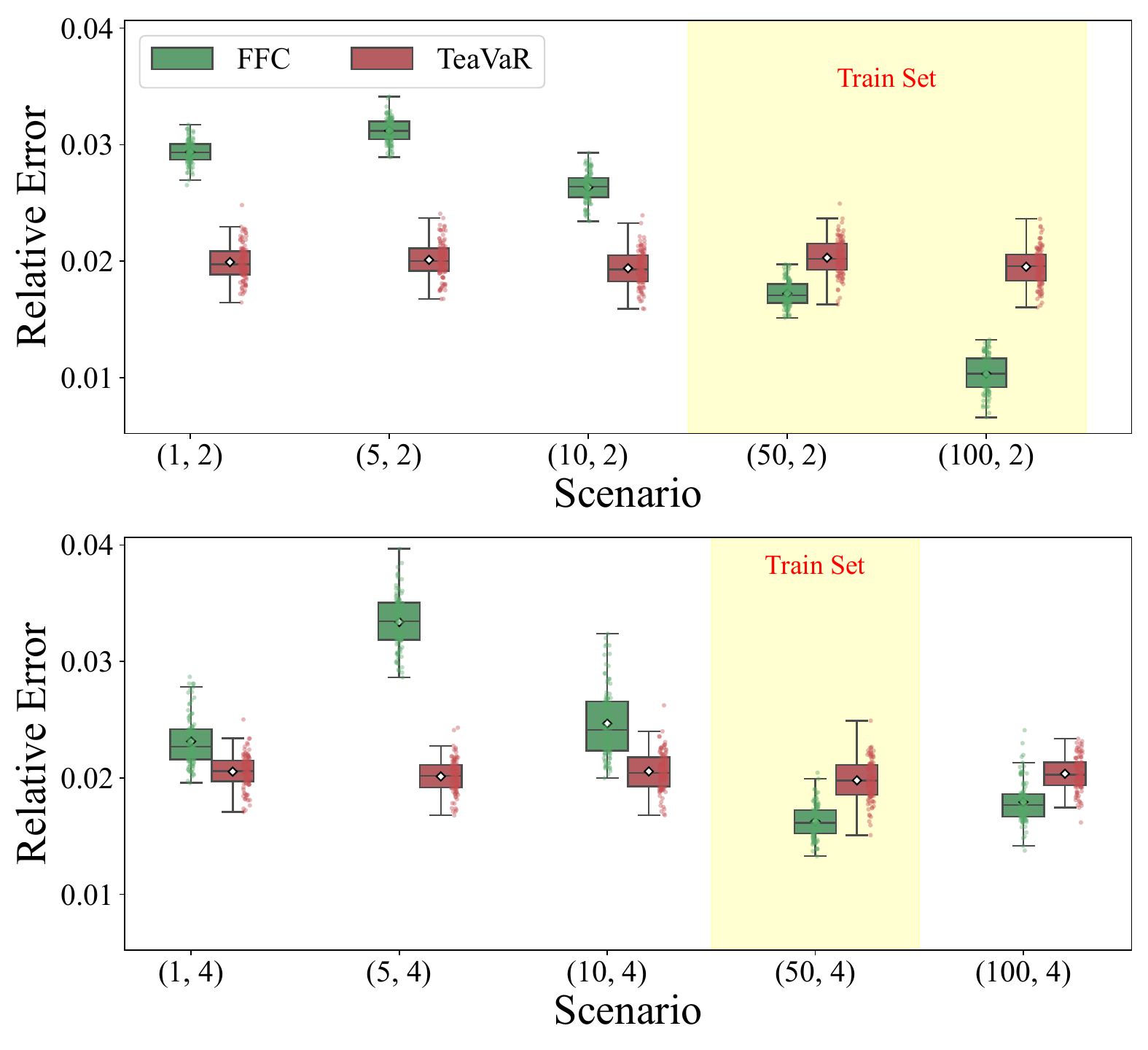}
\caption{\textbf{Scenario-distribution generalization on \textsc{GEANT}.} Same as above.}
\label{fig:app_geant_relerr}
\end{figure}
Figures~\ref{fig:app_b4_relerr}--\ref{fig:app_geant_relerr} report scenario-configuration generalization on \textsc{B4}, \textsc{IBM}, and \textsc{GEANT}, respectively. Each plot evaluates fixed model weights on unseen scenario configurations $(c,s)$ (cutoff thresholds and Weibull scales), and compares the relative error against solver baselines (FFC/PreTE/TeaVaR). Bars marked with yellow background indicate the configurations used during training, while the remaining bars correspond to held-out test configurations. These full plots are the uncropped counterparts of the view shown in Figure~\ref{fig:generalization_summary}(a).
\begin{figure}[thb]
\centering
\includegraphics[width=\columnwidth]{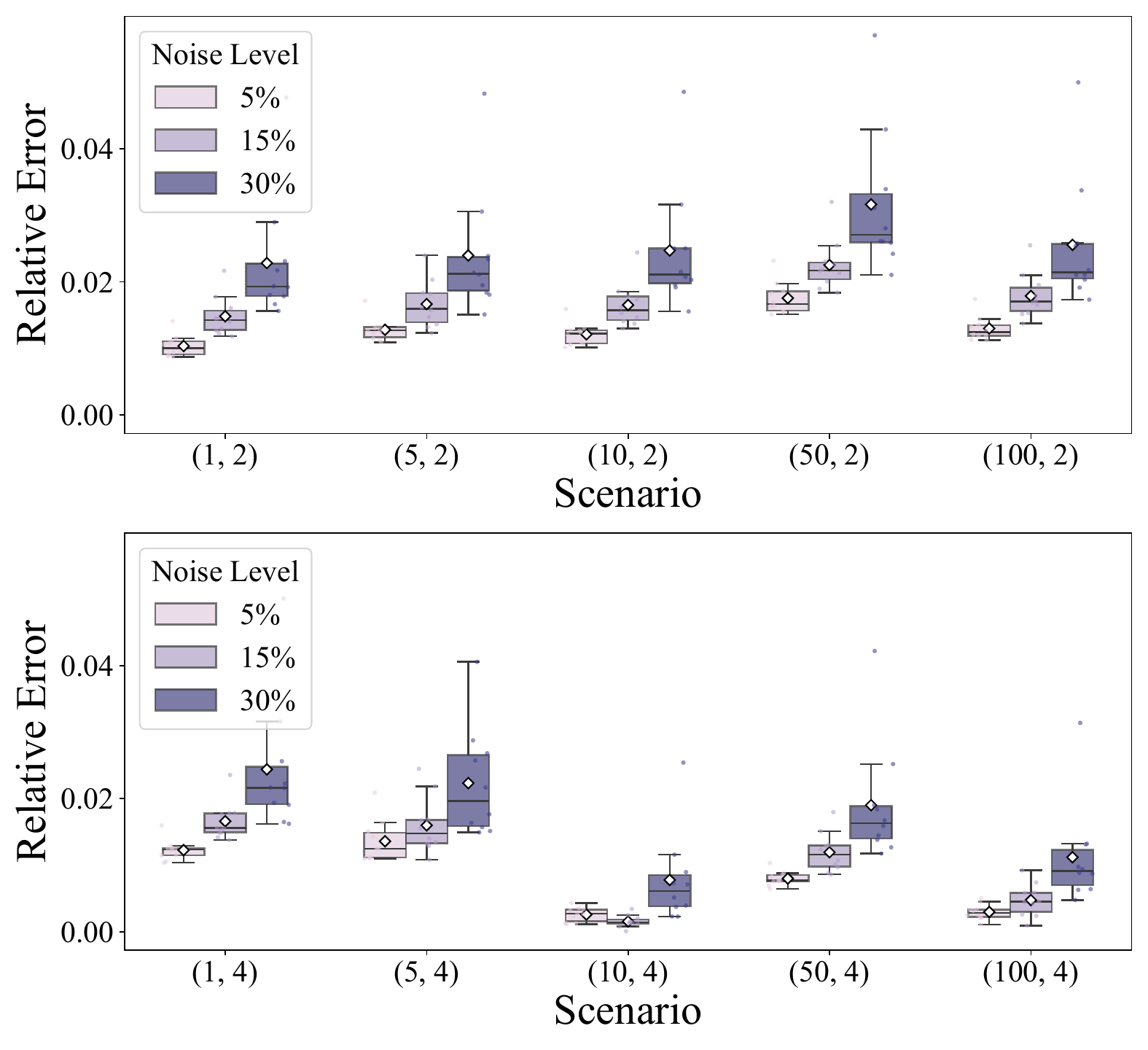}
\caption{\textbf{Demand-perturbation robustness on \textsc{B4} (TeaVaR).} Relative error under demand-proportional Gaussian noise with $\sigma\in\{5\%,15\%,30\%\}$; demands are clipped to be nonnegative.}
\label{fig:app_b4_teavarnoise}
\end{figure}
\begin{figure}[thb]
\centering
\includegraphics[width=\columnwidth]{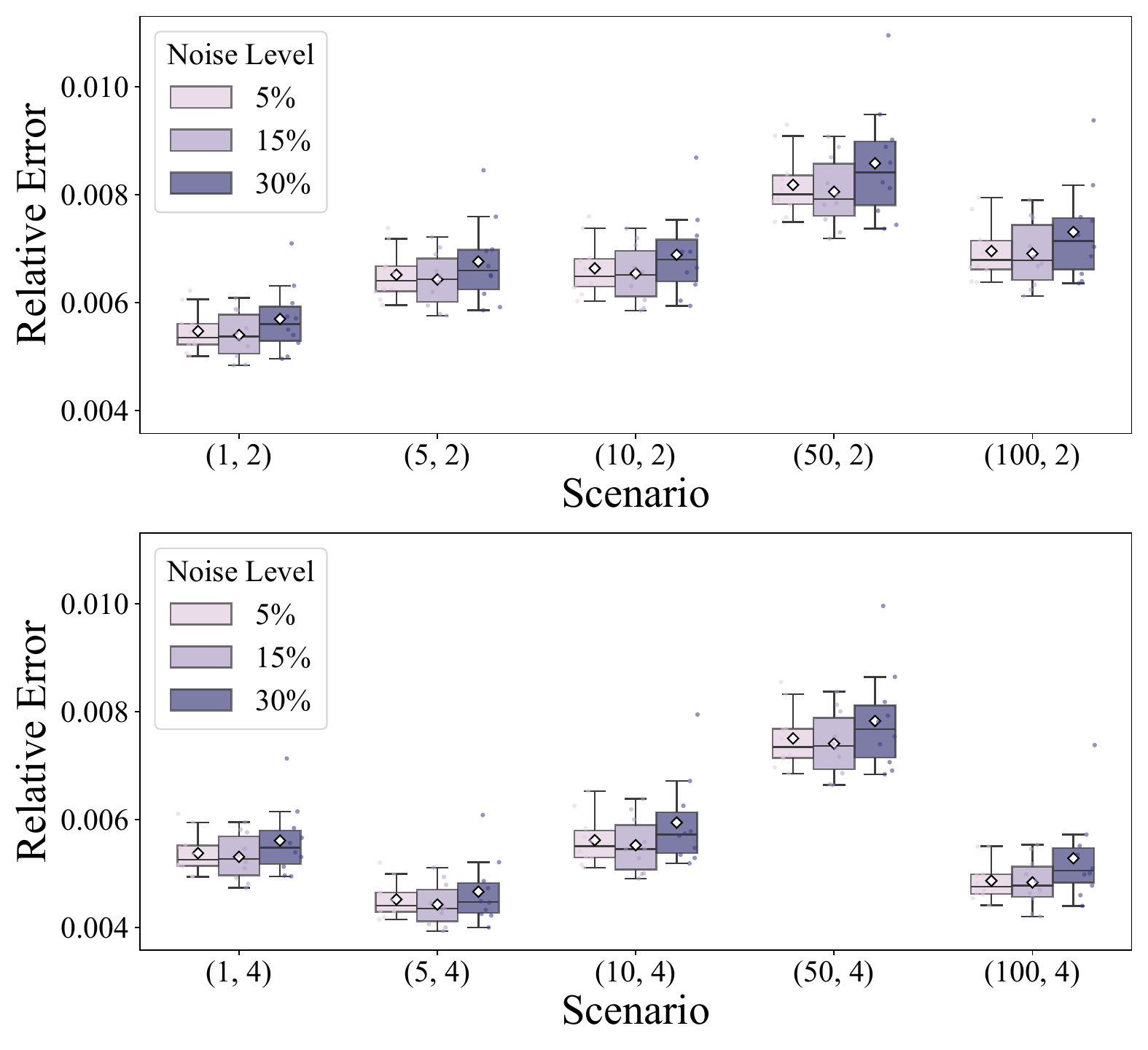}
\caption{\textbf{Demand-perturbation robustness on \textsc{B4} (FFC).} Same as above.}
\label{fig:app_b4_ffcnoise}
\end{figure}
\begin{figure}[thb]
\centering
\includegraphics[width=\columnwidth]{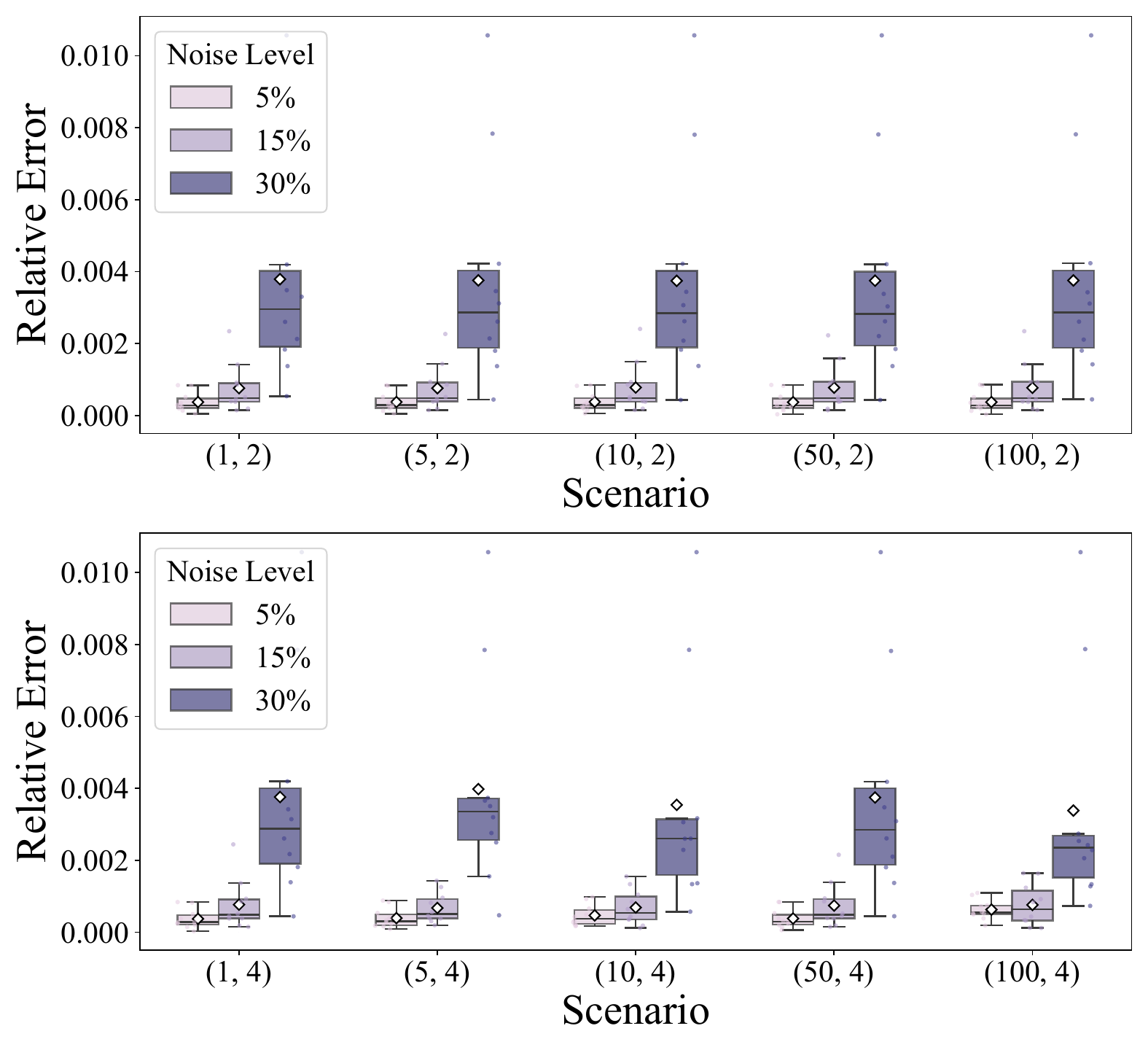}
\caption{\textbf{Demand-perturbation robustness on \textsc{B4} (PreTE).} Same as above.}
\label{fig:app_b4_pretenoise}
\end{figure}
Figures~\ref{fig:app_b4_teavarnoise}--\ref{fig:app_b4_pretenoise} provide the complete demand-perturbation results on \textsc{B4} for TeaVaR, FFC, and PreTE objectives, respectively. Across objectives, we apply the same perturbation protocol described in \S\ref{subsec:eval_noise}: we inject zero-mean Gaussian noise with standard deviation proportional to each demand entry and clip negative demands to zero, while keeping the training set and model checkpoint fixed. These figures extend the representative TeaVaR-only view shown in Figure~\ref{fig:generalization_summary}(b) and illustrate that the robustness trend is consistent across different risk preferences.
\begin{figure}[thb]
\centering
\includegraphics[width=\columnwidth]{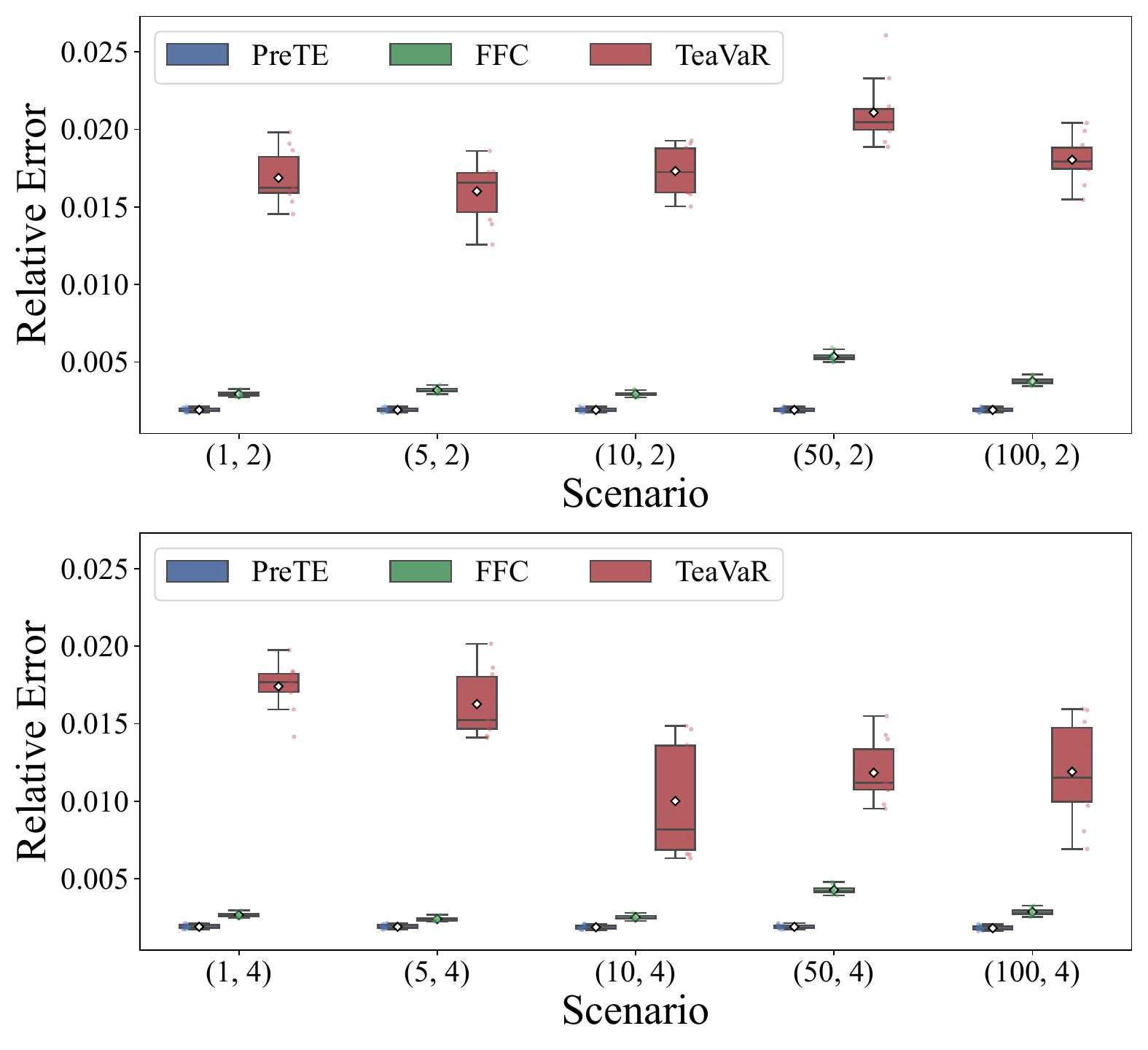}
\caption{\textbf{Tunnel-set generalization on \textsc{B4}.} Relative error under candidate tunnel-set shifts (cross-$K_{sp}$ KSP), evaluated with fixed model weights.}
\label{fig:app_b4_tunnel}
\end{figure}
Figure~\ref{fig:app_b4_tunnel} reports the full version of the tunnel-set shift experiment corresponding to Figure~\ref{fig:generalization_summary}(c). It evaluates the same trained model under altered candidate tunnel sets (generated with a different KSP parameter), demonstrating that the BR reservation parameterization remains compatible with changes in the action space. Since each edge performs a Softmax over the set of tunnels that traverse it, adding or removing tunnels only changes the local Softmax support without requiring architectural modifications.

\section{Iteration Count Generalization}
\label{app:iter_sensitivity}

Table~\ref{tab:iter_sensitivity} provides the full iteration-count generalization results referenced in \S\ref{subsec:eval_iter_generalization}. We deploy a model trained at $K{=}7$ unroll iterations with varying $K_1 \in \{1,2,3,5,7\}$ on \textsc{B4} (TeaVaR objective) without retraining.

\begin{table}[htbp]
\centering
\small
\begin{tabular}{@{}lccccc@{}}
\toprule
$K_1$ & 1 & 2 & 3 & 5 & 7 \\
\midrule
Rel.\ Error (\%) & 7.72 & 4.53 & 3.18 & 1.74 & 1.35 \\
\bottomrule
\end{tabular}
\caption{\textbf{Iteration count generalization on \textsc{B4} (TeaVaR).} A model trained at $K{=}7$ is deployed at $K_1 \le K$ without retraining. Relative error degrades gracefully as $K_1$ decreases, enabling operators to trade quality for speed at runtime.}
\label{tab:iter_sensitivity}
\end{table}

\section{Additional Ablation Diagnostics}
\label{app:ablation_diagnostics}

This appendix provides supplementary diagnostics for the ablation study in \S\ref{subsec:ablation}, focusing on two aspects: (1) route structure under risk-aware objectives, and (2) performance in the nominal (no-failure) setting.

\noindent\textbf{Route structure under TeaVaR objective.}
Table~\ref{tab:teavar_route_structure} reports the fraction of carried traffic allocated to direct (1-hop) tunnels versus multi-hop tunnels on \textsc{B4}.
LS exhibits the highest direct ratio (0.989), confirming that LS induces a strong short-tunnel bias: nearly all traffic is routed via direct tunnels, leaving multi-hop capacity underutilized.
In contrast, GR achieves a lower direct ratio (0.883), indicating better exploitation of tunnel diversity.

\begin{table}[tbh]
\centering
\small
\begin{tabular}{@{}lcccc@{}}
\toprule
Variant & Input dim & Direct & Multi-hop \\
\midrule
GR & 8 & 0.883 & 0.117 \\
BR & 8 & 0.924 & 0.076 \\
LS& 5 & 0.989 & 0.011 \\
GS & 5 & 0.903 & 0.097 \\
\bottomrule
\end{tabular}
\caption{\textbf{Route structure in TeaVaR ablation on \textsc{B4} (mean).} Direct ratio = fraction of \emph{carried traffic} allocated to direct (1-hop) tunnels; multi-hop ratio = $1-\text{direct ratio}$.}
\label{tab:teavar_route_structure}
\end{table}

\noindent\textbf{Nominal (no-failure) throughput.}
Table~\ref{tab:maxflow_nominal} evaluates the same methods on a standard maxflow objective without failure scenarios.
GR achieves the highest normalized throughput on both topologies (0.97 on \textsc{GEANT}, 0.96 on \textsc{B4}), demonstrating that its feasibility-by-construction design is effective for general TE, not only for risk-aware objectives.
GS performs poorly (0.13 on \textsc{GEANT}, 0.17 on \textsc{B4}) because global scaling over-penalizes all tunnels when any single edge is congested (\S\ref{sec:challenges}).
LS and BR achieve intermediate throughput, with LS outperforming BR on \textsc{B4} and \textsc{GEANT}.
Notably, GR and LS have similar direct ratios in the nominal setting (0.40--0.42 on \textsc{B4}), yet GR achieves higher throughput, suggesting that GR's advantage stems from better \emph{allocation} within tunnels rather than tunnel selection alone.

\begin{table}[tbh]
\centering
\small
\begin{tabular}{@{}lccccc@{}}
\toprule
& \multicolumn{2}{c}{\textsc{GEANT}} & \multicolumn{2}{c}{\textsc{B4}} \\
\cmidrule(lr){2-3} \cmidrule(lr){4-5}
Method & Direct & Thr. & Direct & Thr. \\
\midrule
LS & 0.31 & 0.64 & 0.40 & 0.88 \\
GR (Ours) & 0.42 & 0.97 & 0.40 & 0.96 \\
GS & 0.11 & 0.13 & 0.12 & 0.17 \\
BR & 0.29 & 0.47 & 0.31 & 0.70 \\
\bottomrule
\end{tabular}
\caption{\textbf{Nominal maxflow and direct ratio (mean).} Direct ratio = fraction of \emph{carried traffic} allocated to direct (1-hop) tunnels; norm.\ throughput = delivered vs.\ demand (no-failure).}
\label{tab:maxflow_nominal}
\end{table}

\section{Training Cost}
\label{app:training_convergence}

\begin{figure}[tbh]
\centering
\includegraphics[width=\linewidth]{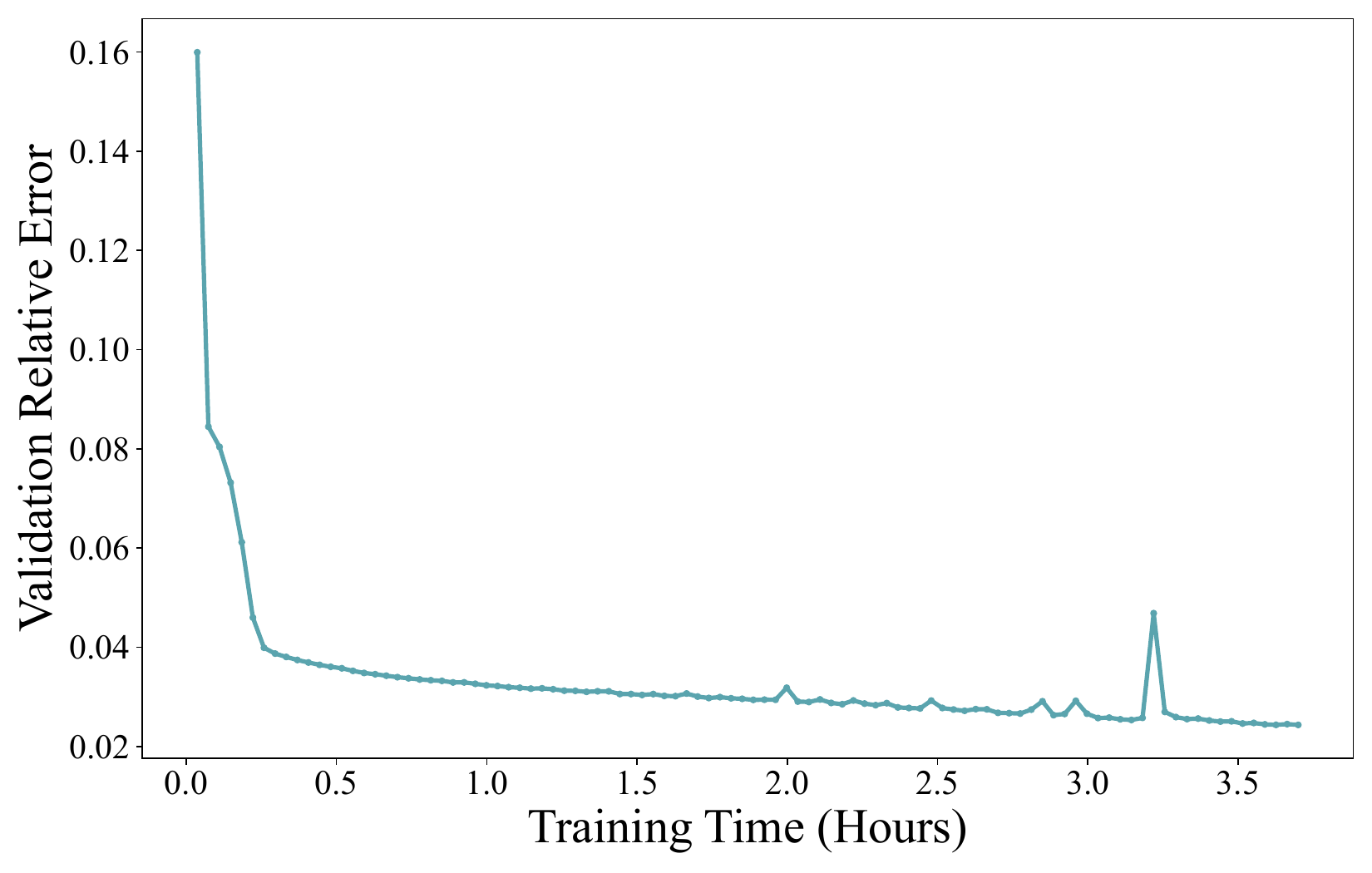}
\caption{\textbf{Training convergence on \textsc{GERMANY50}.} Validation relative error (blue, left axis) and training objective (red, right axis) versus wall-clock time. Scenario chunking (chunk size 20) and gradient checkpointing are both enabled. Training converges within ${\sim}$4\,hours on a single GPU.}
\label{fig:training_convergence}
\end{figure}

Table~\ref{tab:training_time} reports the wall-clock training time for each topology on a single NVIDIA H200 GPU.
Small topologies (\textsc{B4}, \textsc{IBM}) converge in under 10 minutes; \textsc{GEANT} finishes within 15 minutes; the largest topology \textsc{GERMANY50} (with scenario chunking at chunk size 20 and gradient checkpointing both enabled) converges in under 4 hours.
Figure~\ref{fig:training_convergence} shows the convergence curve for this worst-case configuration: the relative error drops from ${\sim}$0.16 to ${\sim}$0.04 in the first 20 minutes, with steady refinement thereafter; the training objective and validation error track closely, indicating no overfitting.

\begin{table}[tbh]
\centering
\small
\begin{tabular}{@{}lrl@{}}
\toprule
Topology & Training time & Note \\
\midrule
\textsc{B4} & $<$\,10\,min & \\
\textsc{IBM} & $<$\,10\,min & \\
\textsc{GEANT} & $\sim$\,15\,min & \\
\textsc{GERMANY50} & $\sim$\,4\,h & chunk\,=\,20, checkpointing \\
\bottomrule
\end{tabular}
\caption{\textbf{One-time training cost per topology} on a single NVIDIA H200 GPU. Training is offline and amortized over all subsequent demand-scenario instances.}
\label{tab:training_time}
\end{table}

